\documentclass[onecolumn]{aa}
\usepackage{amsmath}
\usepackage{amssymb}
\usepackage{graphicx}
\usepackage{wasysym}

\newcommand{\deb}[2]{\displaystyle \frac{\mathrm{d} #1}{\mathrm{d} #2}}

\newcommand{\pdb}[2]{\displaystyle \frac{\partial #1}{\partial #2}}

\newcommand{\dd}[1]{\mathrm{d}^{#1}}
\newcommand{\grad}[1]{\vf{\nabla}#1}
\newcommand{\vf}[1]{{\bf{#1}}}
\newcommand{\vfg}[1]{{\boldsymbol{#1}}}
\newcommand{\ten}[1]{\boldsymbol{\mathsf{#1}}}
\newcommand{\im}{\mathrm{i}}
\newcommand{\omt}{\widetilde{\omega}}
\newcommand{\Btheta}{B_{\vartheta}}
\newcommand{\Bphi}{B_{\varphi}}
\newcommand{\operator}[1]{\mathcal{#1}}
\newcommand{\ppsi}{\partial_{\psi}}
\newcommand{\ptheta}{\partial_{\vartheta}}
\newcommand{\D}{\operator{D}}
\newcommand{\Dd}{\operator{D}^{\dagger}}
\newcommand{\dD}{^\dagger\operator{D}}
\newcommand{\F}{\operator{F}}
\newcommand{\G}{\operator{G}}
\newcommand{\kp}{\kappa_{\mathrm{p}}}
\newcommand{\kt}{\kappa_{\mathrm{t}}}
\newcommand{\kg}{\kappa_{\mathrm{g}}}
\newcommand{\matel}[1]{\mathrm{#1}}
\newcommand{\ephi}{\vf{e}_{\varphi}}

\begin{document}

\title{Unstable magnetohydrodynamical continuous spectrum of accretion disks}
\subtitle{A new route to magnetohydrodynamical turbulence in accretion disks}
\author{J.W.S. Blokland\inst{1} \and R. Keppens\inst{1,2,3} \and J.P. Goedbloed\inst{1,3}}
\institute{FOM-Institute for Plasma Physics 'Rijnhuizen', 
           Association Euratom-FOM, Trilateral Euregio Cluster, 
	   P.O. Box 1207, 3430 BE Nieuwegein, The Netherlands
           \and
	   Centre for Plasma Astrophysics, K.U. Leuven, Belgium
	   \and
	   Astronomical Institute, Utrecht University, The Netherlands}
\offprints{J.W.S. Blokland \email{J-W.S.Blokland@Rijnhuizen.nl}}
\date{Received / Accepted}
\abstract{% context heading (optional)
          We present a detailed study of localised magnetohydrodynamical (MHD) instabilities occuring in
          two--dimensional magnetized accretion disks.
         }
	 {% aims headings (mandatory)
	  We model axisymmetric MHD disk tori, and solve the equations governing a two--dimensional magnetized
	  accretion disk equilibrium and linear wave modes about this equilibrium. We show the existence of 
	  novel MHD instabilities in these two--dimensional equilibria which do not occur in an accretion disk in 
	  the cylindrical limit.
	 }
	 {% methods heading (mandatory)
	  The disk equilibria are numerically computed by the FINESSE code. The stability of accretion disks is 
	  investigated analytically as well as numerically. We use the PHOENIX code to compute all the waves and
	  instabilities accessible to the computed disk equilibrium.
	 }
	 {% results heading (mandatory)
	  We concentrate on strongly magnetized disks and sub--Keplerian rotation in a large part of the disk. 
	  These disk equilibria show that the thermal pressure of the disk can only decrease outwards if there is 
	  a strong gravitational potential. Our theoretical stability analysis shows that convective continuum
	  instabilities can only appear if the density contours coincide with the poloidal magnetic flux contours. 
	  Our numerical results confirm and complement this theoretical analysis. Furthermore, these results show 
	  that the influence of gravity can either be stabilizing or destabilizing on this new kind of MHD instability. 
	  In the likely case of a non--constant density, the height of the disk should exceed a threshold before this 
	  type of instability can play a role.
	 }
	 {% conclusions heading (optional)
	  This localised MHD instability provides an ideal, linear route to MHD turbulence in strongly magnetized
	  accretion disk tori.
	 }
\titlerunning{Unstable MHD continuous spectrum of accretion disks}
\authorrunning{J. W. S. Blokland et al.}
\maketitle

\section{Introduction}
Accretion disks are very common objects in astrophysics. These objects can be found in stellar systems as well as in
active galactic nuclei (AGN). In the first case, the disk accretes matter onto a protostar
(young stellar object (YSO)), white dwarf, neutron star, or a black hole. The typical size of these disks is a few 
hundred astronomical units (AU). A massive black hole is the central object in the case of an accretion disk in AGN.
Here, the typical size of the disk is a hundred parsecs (pc). 

Much research on accretion disk dynamics focuses on occuring accretion processes in an MHD framework. This is done
linearly as well as non--linearly. The linear studies aim to understand the drivers of the accretion mechanism. 
This can be done by looking at waves and instabilities about a disk equilibrium in the cylindrical limit 
(see e.g. Keppens et al.~\cite{KCG}, Blokland et al.~\cite{BSKG}, van der Swaluw et al.~\cite{SBK}, and references
therein). In this limit the disk equilibrium is essentially one--dimensional and one can use a self--similar model 
like the one of Spruit et al.~(\cite{SMIS}).  When instabilities are found, one must compute the resulting evolution 
non--linearly to see if these instabilities give rise to turbulence. This turbulence may be the source of an enhanced 
effective viscosity mechanism which causes an increased transport of angular momentum outwards, as  needed for accretion 
(Shakura \& Sunyaev~\cite{ShSu}). Balbus \& Hawley (\cite{BaHa}) discussed that the MHD turbulence resulting from 
the magneto--rotational instability (Velikhov~\cite{Ve} and Chandrasekhar~\cite{Ch}) could provide this angular 
momentum transport.

In the last years, global two or even three--dimensional magnetohydrodynamical (MHD) simulations of accretion disks have
been performed (see for example Matsumoto \& Shibata~\cite{MS} and Armitage~\cite{Am}). In these simulations, the 
non--linear dynamics is usually interpreted as a direct consequence of the magneto--rotational instability 
(see for example Hawley et al.~\cite{HBS}) leading to angular momentum transport outwards. Another possibility for the 
initial and evolving dynamics is that both the convective and the magneto--rotational instability play an important 
role in the angular momentum transport (see for example Igumenshchev et al.~\cite{INA}). In the latter case, these disks 
are known as convection--dominated accretion flows (CDAF). However,especially those
global simulations starting from initial axisymmetric accretion tori (see for example Hawley~\cite{Haw}) are at best 
loosely connected to the linear stability studies. A detailed catalogue of the MHD spectrum of such two--dimensional 
disks is completely lacking. Moreover, the usual identification of the cause of the turbulent dynamics with the linear
magneto--rotational or convective instability is mostly based on the extrapolation of the linear stability analysis 
from a one--dimensional to a two or three--dimensional accretion disk. Such extrapolating ignores the fact that if one 
performs a detailed MHD stability study of a two--dimensional accretion disk, one may find many new types of 
instability that all could lead to effective angular momentum transport. A recent example of such new, poloidal 
flow--driven type of instability is presented by Goedbloed et al.~(\cite{GBHK}), where the Trans--slow Alfv\'en 
Continuum (TSAC) mode is shown to occur inside a disk with toroidal and super--slow poloidal flow in the presence 
of a strong gravitational potential.

This paper has two aims. The first one is to present the equations and the numerical solutions of two--dimensional MHD 
disk equilibria with toroidal flows, but without poloidal flows. This case is fundamentally different from the previously
discussed one of combined poloidal and toroidal flows (Goedbloed et al.~\cite{GBHK}) because the equilibria are described
by quite different flux functions. By considering an axisymmetric equilibrium, we are able to model a thick 
accretion disk without any further approximations. All the equilibria presented below are computed using the code
FINESSE~(Beli\"en et al. \cite{BBGHK}). The second aim is to present a detailed stability analysis of these
two--dimensional equilibria. The analysis is done theoretically as well as numerically. For the stability computations
we have used the recently published spectral code~PHOENIX~(Blokland et al.~\cite{BHKG}). To our knowlegde, this kind 
of analysis for MHD disk equilibria with purely toroidal flow has never been presented in the
astrophysical literature until now.

We will present a sample of disk equilibria where the disk plasma typically varies from strongly magnetized up to close
to equipartition. Also, the rotational speed of the plasma varies from sub--Keplerian up to Keplerian. The theoretical
part of the spectral analysis of these equilibria shows that the convective continuum instabilities may appear inside the
disk. These instabilities are part of the continuous branches that exist in the MHD spectrum of the linear eigenmodes
of the disk tori and are localised on magnetic flux surfaces. We derive a stability criterion for this instability. 
This criterion looks similar to the Schwarzschild criterion. However, our criterion governs convective instabilities
along the poloidal magnetic field lines while the Schwarzschild criterion applies in the direction perpendicular
to the poloidal magnetic field lines. This theoretical analysis has been verified by our numerical stability calculations.

The paper is organised as follows. In Sect.~\ref{sec:equilibrium}, we present the equations which govern a
two--dimensional accretion disk equilibrium. In Sect.~\ref{sec:spectral}, we recall the essential elements from 
spectral theory of MHD waves and instabilities, such as the Frieman and Rotenberg formalism 
(Frieman \& Rotenberg~\cite{FR}), straight field line coordinates and representation. In Sect.~\ref{sec:continuous},
these are used to derive the equations for the continuous MHD spectrum  and the stability criterion for the convective
instability. The numerical codes FINESSE and PHOENIX are discussed in Sect.~\ref{sec:codes}. In Sect.~\ref{sec:results},
we present our numerical results on the disk equilibria and their stability analysis. Finally, in
Sect.~\ref{sec:conclusions}, we summarise and present our conclusions.

\section{Accretion disk equilibrium \label{sec:equilibrium}}
We consider an axisymmetric accretion disk. Because of this symmetry, cylindrical coordinates $(R,Z,\varphi)$ are 
used and the equilibrium quantities only depend on the poloidal coordinates~$R$ and $Z$.  The disk equilibrium is 
modeled by the ideal MHD equations,
\begin{align}
  \label{eq:momentum}
  \rho\pdb{\vf{v}}{t} & = -\rho\vf{v}\cdot\grad{\vf{v}} - \grad{p} + \vf{j}\times\vf{B} - \rho\grad{\Phi}, \\
  \label{eq:entropy}
  \pdb{p}{t}          & = -\vf{v}\cdot\grad{p} - \gamma p\grad{}\cdot\vf{v}                       , \\
  \label{eq:induction}
  \pdb{\vf{B}}{t}     & =  \grad{}\times\left(\vf{v}\times\vf{B}\right)                           , \\
  \label{eq:mass}
  \pdb{\rho}{t}       & = -\grad{}\cdot\left(\rho\vf{v}\right)                                    ,
\end{align}
where $\rho$, $p$, $\vf{v}$, $\vf{B}$, $\Phi$ and $\gamma$ are the density, pressure, velocity, magnetic field, 
gravitational potential, and the ratio of the specific heats, respectively.  The current 
density~$\vf{j} = \grad{}\times\vf{B}$ and the equation $\grad{}\cdot\vf{B}=0$ has to be satisfied. 
Furthermore, the disk equilibrium is assumed to be time-independent. The relation between the density and the 
thermal pressure is expressed by the ideal gas law:~$p = \rho T$. From the equations~$\grad{}\cdot\vf{B}=0$ 
and~$\grad{}\cdot\vf{j}=0$ it follows that the magnetic field and the current density can be written as
\begin{align}
  \vf{B} & =  \frac{1}{R}\ephi\times\grad{\psi} + \Bphi\ephi,        \\
  \vf{j} & = -\frac{1}{R}\ephi\times\grad{I}    + j_{\varphi}\ephi,
\end{align}
respectively. Here, $2\pi\psi$ is the poloidal flux and the poloidal stream function~$I = R\Bphi$. We restrict ourselves
to disk equilibria with purely toroidal flow,
\begin{equation}
  \vf{v} = v_{\varphi}\ephi = R\Omega\ephi.
  \label{eq:velocity}
\end{equation}
In this case, equations~\eqref{eq:entropy} and~\eqref{eq:mass} are trivially satisfied. The angular 
velocity~$\Omega$ is related to the electric field,
\begin{equation}
  \vf{E} = -\vf{v}\times\vf{B} = \Omega\grad{\psi},
\end{equation}
and making use of the induction equation~\eqref{eq:induction} it can be easily shown that $\Omega = \Omega(\psi)$.

Three different projections can be applied on the momentum equation~\eqref{eq:momentum}. The projections are in the 
toroidal direction and in the poloidal plane parallel and perpendicular to the poloidal magnetic field lines. From 
the toroidal projection one can show that the poloidal stream function is a flux function, i.e.~$I = I(\psi)$.
The projection parallel to the poloidal magnetic field results in two equations,
\begin{equation}
  \begin{aligned}
    \left. \pdb{p}{R} \right|_{\psi=\mathrm{const}} & =  \rho\left( R\Omega^{2} - \pdb{\Phi}{R} \right),      \\
    \left. \pdb{p}{Z} \right|_{\psi=\mathrm{const}} & = -\rho                     \pdb{\Phi}{Z}        ,
  \end{aligned}
  \label{eq:parallelBpol}
\end{equation}
which have to be satisfied simultaneously. From these two equations we conclude that the pressure~$p = p(\psi; R, Z)$.
The last projection, perpendicular to the poloidal magnetic field, leads to the extended Grad-Shafranov equation,
\begin{equation}
  R^{2}\grad{}\cdot\left( \frac{1}{R^{2}}\grad{\psi} \right) = - I\deb{I}{\psi} - R^{2}\pdb{p}{\psi}.
  \label{eq:gradshafranov}
\end{equation}
The two equations parallel to the poloidal magnetic field~\eqref{eq:parallelBpol} can be solved analytically under the
extra assumption that either the temperature~$T$, or the density~$\rho$, or the entropy~$S=p\rho^{-\gamma}$ is a 
flux function. The assumption that the temperature is a flux function can be justified due to the high thermal 
conductivity along the magnetic field lines. This is true at least up to the transport time scale, which is long 
compared to the Alfv\'en time. The resulting pressure can then be written as
\begin{equation}
  p(\psi;R,Z) = p_{0}(\psi) \exp \left[ (R^{2} - R_{0}^{2}) \Lambda_{T}(\psi) - \frac{\Phi(R,Z)}{T(\psi)} \right],
  \label{eq:pressureT}
\end{equation}
where $\Lambda_{T} \equiv \Omega^{2} / (2T)$ is a flux function, $p_{0}$ is the pressure for a static pure Grad--Shafranov
equilibrium without gravity, and $R_{0}$ is the geometric axis of the accretion disk. The extended Grad-Shafranov 
equation~\eqref{eq:gradshafranov} reduces to
\begin{equation}
  \begin{split}
    R^{2}\grad{}\cdot\left( \frac{1}{R^{2}}\grad{\psi} \right) & = 
      -I\deb{I}{\psi} - R^{2}\left\{ \deb{p_{0}}{\psi} + p_{0}\left[ (R^{2}-R_{0}^{2})\deb{\Lambda_{T}}{\psi} + \frac{\Phi}{T^{2}}\deb{T}{\psi} \right] \right\} \\
      & \phantom{= -I\deb{I}{\psi} - R^{2}}
        \times \exp \left[ \left( R^{2} - R_{0}^{2} \right) \Lambda_{T} - \frac{\Phi}{T} \right].
  \end{split}
  \label{eq:gradshafranovT}
\end{equation}

Another possibility, on MHD time scales, is to assume that the density is a flux function. In this case, the pressure
reads
\begin{equation}
  p(\psi;R,Z) = p_{0}(\psi) \left[ 1 + (R^{2}-R_{0}^{2})\Lambda_{\rho}(\psi) - \frac{\Phi(R,Z)}{T_{\rho}(\psi)} \right],
  \label{eq:pressurerho}
\end{equation}
where~ the quasi-temperature~$T_{\rho} \equiv p_{0} / \rho$ and $\Lambda_{\rho} \equiv \Omega^{2} / (2T_{\rho})$. 
Under this assumption, the extended Grad-Shafranov equation~\eqref{eq:gradshafranov} can be written as
\begin{equation}
  \begin{split}
    R^{2}\grad{}\cdot\left( \frac{1}{R^{2}}\grad{\psi} \right) & =
      -I\deb{I}{\psi} - R^{2} \left\{ \deb{p_{0}}{\psi} + p_{0} \left[ (R^{2}-R_{0}^{2})\deb{\Lambda_{\rho}}{\psi} + \frac{\Phi}{T_{\rho}^{2}}\deb{T_{\rho}}{\psi} \right] \right.
      \left. \left[ 1 + (R^{2}-R_{0}^{2})\Lambda_{\rho} - \frac{\Phi}{T_{\rho}} \right]^{-1} \right\} \\
    & \phantom{= -I\deb{I}{\psi} - R^{2}}
      \times \left[ 1 + (R^{2}-R_{0}^{2})\Lambda_{\rho} - \frac{\Phi}{T_{\rho}} \right].
  \end{split}
  \label{eq:gradshafranovrho}
\end{equation}

The final option is to assume that the entropy~$S$ is a flux function. The advantage of this assumption is that it
allows for a natural extension to include both toroidal and poloidol flows in the equilibrium, where the entropy has
to be a flux function (Zehrfeld et al. \cite{ZG}; Hameiri \cite{Ham}). In this case, the pressure reads
\begin{equation}
  p(\psi;R,Z) = p_{0}(\psi) \left\{ 1 + \frac{\gamma - 1}{\gamma} \left[ (R^{2}-R_{0}^{2})\Lambda_{S}(\psi) - \frac{\Phi(R,Z)}{T_{S}(\psi)} \right] \right\}^{\gamma/(\gamma - 1)},
  \label{eq:pressureS}
\end{equation}
and the extended Grad-Shafranov equation~\eqref{eq:gradshafranov} reduces to
\begin{equation}
  \begin{split}
    R^{2}\grad{}\cdot\left( \frac{1}{R^{2}} \grad{\psi} \right) & =
      -I\deb{I}{\psi} - R^{2} \left\{ \deb{p_{0}}{\psi} + p_{0} \left[ (R^{2}-R_{0}^{2})\deb{\Lambda_{S}}{\psi} + \frac{\Phi}{T_{S}^{2}}\deb{T_{S}}{\psi} \right] \right.
      \left. \left[ 1 + \frac{\gamma - 1}{\gamma} \left( (R^{2}-R_{0}^{2})\Lambda_{S} - \frac{\Phi}{T_{S}} \right) \right]^{-1} \right\} \\
    & \phantom{= -I\deb{I}{\psi} - R^{2}}
      \times \left\{ 1 + \frac{\gamma - 1}{\gamma} \left[ (R^{2}-R_{0}^{2})\Lambda_{S} - \frac{\Phi}{T_{S}} \right] \right\}^{\gamma / (\gamma-1)},
  \end{split}
  \label{eq:gradshafranovS}
\end{equation}
where the quasi-temperature~$T_{S} \equiv S\rho_{0}^{\gamma-1}$ and $\Lambda_{S} \equiv \Omega^{2} / (2T_{S})$. 
The extended Grad-Shafranov equation has been used previously by van der Holst et al.~(\cite{HBG}) for the first
two cases without gravity.

The density can be easily derived for all three assumptions by inserting the corresponding equation for the pressure
into the momentum equations parallel to the poloidal magnetic field lines~\eqref{eq:parallelBpol}. The resulting
density is
\begin{equation}
  \rho(\psi; R,Z) = \rho_{0}(\psi) \times
  \begin{cases}
    \left[ \left( R^{2} - R_{0}^{2} \right) \Lambda_{T} - \dfrac{\Phi}{T}  \right]  &
      \text{for }T=T(\psi)                                                          \\
    1                                                                               &
      \text{for }\rho=\rho(\psi)                                                    \\
    \left\{ 1 + \dfrac{\gamma -1}{\gamma} \left[ \left( R^{2} - R_{0}^{2} \right) \Lambda_{S} - \dfrac{\Phi}{T_{S}} \right] \right\}^{1/(\gamma - 1)}  &
      \text{for }S=S(\psi)
  \end{cases}.
\end{equation}
The flux function~$\rho_{0}$ corresponds to the density of a static equilibrium without gravity. All these cases with the 
inclusion of gravity have been discussed in the appendix to Blokland et al.~(\cite{BHKG}).

\section{Spectral formulation \label{sec:spectral}}
\subsection{Frieman--Rotenberg formalism}
For the investigation of stability properties of stationary MHD equilibria, the formalism by Frieman and 
Rotenberg~(\cite{FR}) has been used. They derived from the linearised MHD equations, one second order differential 
equation for the Lagrangian displacement vector field~$\vfg{\xi}$,
\begin{equation}
  \vf{F}(\vfg{\xi}) - 2\rho\vf{v}\cdot\grad{\pdb{\vfg{\xi}}{t}} - \rho\frac{\partial^{2}\vfg{\xi}}{\partial t^{2}} = 0,
  \label{eq:friemanrotenberg}
\end{equation}
where the force operator $\vf{F}(\vfg{\xi})$ is
\begin{align}
  \label{eq:forceoperator}
  \vf{F}(\vfg{\xi})                   & = 
    \vf{F}_{\mathrm{static}}(\vfg{\xi}) + \grad{\Phi}\grad{}\cdot(\rho\vfg{\xi})+ 
    \grad{}\cdot\left[ \rho\vfg{\xi}\vf{v}\cdot\grad{\vf{v}} - \rho\vf{v}\vf{v}\cdot\grad{\vfg{\xi}} \right]. \\
\intertext{Here,}
  \label{eq:forceoperatorstatic}
  \vf{F}_{\mathrm{static}}(\vfg{\xi}) & =
    -\grad{\Pi} + \vf{B}\cdot\grad{\vf{Q}} + \vf{Q}\cdot\grad{\vf{B}}
\end{align}
is the force operator for static equilibria without gravity, which has been derived by Bernstein et al.~(\cite{BFKK}).
Here, the Eulerian perturbation of the total pressure,
\begin{align}
  \label{eq:Etotalpressure}
  \Pi    & = -\gamma p\grad{}\cdot\vfg{\xi} - \vfg{\xi}\cdot\grad{p} + \vf{B}\cdot\vf{Q}, \\
  \intertext{and the Eulerian perturbation of the magnetic field,}
  \label{eq:Emagneticfield}
  \vf{Q} & = \grad{} \times \left( \vfg{\xi} \times \vf{B} \right).
\end{align}

The time--dependence for the displacement field is assumed to be an exponential one with normal mode
frequencies~$\omega$, $\vfg{\xi} = \hat{\vfg{\xi}} \exp ( -\im \omega t )$. Using this assumption, the 
Frieman--Rotenberg equation can be written as
\begin{equation}
  \vf{F}(\vfg{\xi}) + 2\im \rho\omega\vf{v}\cdot\grad{\vfg{\xi}} + \rho\omega^{2}\vfg{\xi} = 0.
  \label{eq:FReqn}
\end{equation}
In the derivations below, the toroidal symmetry of the accretion disk equilibrium is exploited. This is done by 
writing the toroidal dependence of the displacement field as~$\vfg{\xi} \sim \exp (\im n\varphi)$, where~$n$ is the
toroidal mode number.

\subsection{Straight field line coordinates}
The analytical and numerical analysis of MHD waves and instabilities are done in 'straight field line' coordinates.
To convert from cylindrical $(R,Z,\varphi)$ to straight field line coordinates
$(x^{1}\equiv\psi, x^{2}\equiv\vartheta, x^{3}\equiv\varphi)$ one needs the metric tensor and the Jacobian associated
with the non--orthogonal coordinates in which the equilibrium field lines appear straight. This is standard practice
in MHD spectral studies for laboratory tokamak plasmas. The metric elements~$g_{ij}$ and the Jacobian~$J$ are
\begin{equation}
  \begin{aligned}
    g^{ij} & = \grad{x^{i}}                          \cdot \grad{x^{j}},                           & \quad
    g_{ij} & = \frac{\partial\vf{r}}{\partial x^{i}} \cdot \frac{\partial\vf{r}}{\partial x^{j}},  & \quad
    J      & = \left( \grad{\psi}\times\grad{\vartheta} \cdot \grad{\varphi} \right)^{-1},
  \end{aligned}
\end{equation}
respectively. Here, the poloidal angle~$\vartheta$ is constructed such that the magnetic field lines are straight in
the $(\vartheta,\varphi)$--plane. The slope of these lines is a flux function:
\begin{equation}
  \left. \deb{\varphi}{\vartheta} \right|_{\mathrm{field line}} = 
  \frac{\vf{B}\cdot\grad{\varphi}}{\vf{B}\cdot\grad{\vartheta}} = \frac{IJ}{R^{2}} = q(\psi),
  \label{eq:safetyfactor}
\end{equation}
where $q$ is the safety factor. Furthermore, in straight field coordinates the poloidal and toroidal curvature of the 
magnetic surfaces are
\begin{alignat}{2}
  \kp & = -\vf{n} \cdot \left( \vf{t}\cdot\grad{\vf{t}} \right)                     &  
      & = \frac{R}{J}\left( \ppsi - \ptheta\frac{g_{12}}{g_{22}} \right) J\Btheta,  \\
  \kt & = -\vf{n} \cdot \left( \ephi \cdot\grad{\ephi}  \right)                     &  
      & = \Btheta    \left( \ppsi - \frac{g_{12}}{g_{22}}\ptheta \right) R,
\end{alignat}
respectively. Here, $\vf{n} = \grad{\psi} / |\grad{\psi}|$ and $\vf{t} = \vf{B}_{\vartheta} / \Btheta$, where~$\Btheta$
is the poloidal magnetic field. It is important to realise that the straight field coordinates can only be constructed 
when the solution~$\psi(R,Z)$ has been computed from the extended Grad-Shafranov equation~\eqref{eq:gradshafranov}.

\subsection{Field line projection and representation}
On each flux surface the distinction between two wave directions can be made: one parallel and the other one 
perpendicular to the magnetic field. These directions correspond to the short wavelength limit of slow and Alfv\'en 
waves in cylindrical geometry. Hence, it is useful to exploit a projection based on the magnetic surfaces and 
field lines using the triad of unit vectors,
\begin{equation}
  \begin{aligned}
    \vf{n}    & \equiv \frac{\grad{\psi}}{|\grad{\psi}|},  & \quad
    \vfg{\pi} & \equiv \vf{b} \times \vf{n},               & \quad
    \vf{b}    & \equiv \frac{\vf{B}}{B}.
  \end{aligned}
  \label{eq:basevectors}
\end{equation}
Using these unit vectors, the components of the displacement field~$\vfg{\xi}$ can be written as
\begin{equation}
  \begin{aligned}
    X & \equiv     R\Btheta           \vfg{\xi}\cdot\vf{n},     & \quad
    Y & \equiv \im \frac{B}{R\Btheta} \vfg{\xi}\cdot\vfg{\pi},  & \quad
    Z & \equiv \im \frac{1}{B}        \vfg{\xi}\cdot\vf{b},
  \end{aligned}
  \label{eq:xicomponents}
\end{equation}
and the projected gradient operators become
\begin{alignat}{2}
  \label{eq:operatorD}
  \D & \equiv      \frac{1}{R\Btheta}\vf{n}   \cdot\grad{}  &  & = \ppsi - \frac{g_{12}}{g_{22}}\ptheta,  \\
  \label{eq:operatorG}
  \G & \equiv -\im R\Btheta B        \vfg{\pi}\cdot\grad{}  &  & = \frac{-\im I}{J}\ptheta - n\Btheta^{2},  \\
  \label{eq:operatorF}
  \F & \equiv -\im                   \vf{B}   \cdot\grad{}  &  & = \frac{-\im}{J}\ptheta + \frac{nq}{J}.
\end{alignat}
The second equality in the expressions for $\F$ and $\G$ only holds when these operators act on a
component of the displacement field~$\vfg{\xi}$.

Using the straight field line coordinates and the projections~\eqref{eq:xicomponents} the Frieman-Rotenberg 
equation~\eqref{eq:friemanrotenberg} can be written as
\begin{equation}
  \begin{aligned}
    (\ten{A} + \ten{R})\vf{x} + 2\rho\omt\Omega\ten{C}\vf{x} - \rho\omt^{2}\ten{B}\vf{x} & = 0,  & 
    \quad \mathrm{with} \quad
    \vf{x} & \equiv \begin{pmatrix} X \\ Y \\ Z \end{pmatrix},
  \end{aligned}
  \label{eq:spectraleqn}
\end{equation}
where ~$\omt(\psi) \equiv \omega - n\Omega(\psi)$ is the Doppler shifted eigenfrequency. Here,~$\ten{A}$ and~$\ten{B}$ 
are~$3 \times 3$ matrix operators which are also present in the case of a static equilibrium without gravity. The matrix 
operators~$\ten{C}$ and~$\ten{R}$ contain the elements due to the toroidal rotation of the plasma and of an external 
gravitational potential. The matrix elements of~$\ten{A}$ and~$\ten{B}$ are
\begin{equation}
  \begin{aligned}
    \matel{A}_{11} & \equiv -\D\frac{\gamma p + B^{2}}{J}\Dd J + \F\frac{1}{R^{2}B_{\vartheta}^{2}}\F
                            + 2\left[ \D\left(\frac{\Btheta\kp}{R}\right) \right]
	                    + 2\frac{\Bphi}{R}\left[ J \dD\left(\frac{1}{J} \frac{\Bphi\kt}{\Btheta}\right)\right],  \\
    \matel{A}_{12} & \equiv -\D\gamma p\G \frac{1}{B^{2}} - \D\G 
                            + 2\left( \frac{n\Btheta\kp}{R} + \frac{\Bphi\kt}{\Btheta}\frac{\im}{J}\ptheta \right),  \\
    \matel{A}_{13} & \equiv -\D\gamma p\F,                                                                           \\
    \matel{A}_{21} & \equiv \frac{1}{B^{2}}\G\frac{\gamma p}{J}\Dd J + \G\frac{1}{J}\Dd J 
                            + 2\left( \frac{n\Btheta\kp}{R} + \frac{\im}{J}\ptheta\frac{\Bphi\kt}{\Btheta} \right),  \\
    \matel{A}_{22} & \equiv \frac{1}{B^{2}}\G\gamma p\G\frac{1}{B^{2}} + \G\frac{1}{B^{2}}\G 
                            + \F\frac{R^{2}\Btheta^{2}}{B^{2}}\F,                                                    \\
    \matel{A}_{23} & \equiv \frac{1}{B^{2}}\G\gamma p\F,                                                             \\
    \matel{A}_{31} & \equiv \F\frac{\gamma p}{J}\Dd J,                                                               \\
    \matel{A}_{32} & \equiv \F\gamma p\G\frac{1}{B^{2}},                                                             \\
    \matel{A}_{33} & \equiv \F\gamma p\F,                                                                            \\
  \end{aligned}
  \label{eq:matrixA}
\end{equation}
and
\begin{equation}
  \begin{aligned}
    \matel{B}_{11} & \equiv \frac{1}{R^{2}\Btheta^{2}},      &
    \matel{B}_{22} & \equiv \frac{R^{2}\Btheta^{2}}{B^{2}},  &
    \matel{B}_{33} & \equiv B^{2}.
  \end{aligned}
  \label{eq:matrixB}
\end{equation}
Here, the operators
\begin{equation}
  \begin{aligned}
    \Dd & \equiv \D - \left[ \ptheta\left(\frac{g_{12}}{g_{22}}\right) \right],  &
    \dD & \equiv \D + \left[ \ptheta\left(\frac{g_{12}}{g_{22}}\right) \right],
  \end{aligned}
  \label{eq:operatorsD}
\end{equation}
are related to the normal gradient operator~$\D$, which has been introduced by Goedbloed~(\cite{Go-1997}) for the
spectral analysis of static tokamak equilibria.  The square brackets in the expressions~\eqref{eq:matrixA} for the
matrix elements of~$\ten{A}$ and the gradient operators defined by~\eqref{eq:operatorsD} indicate that the
differential operator only acts on the term inside the bracket. This notation is also used in the expressions below. 
The matrices~$\ten{A}$ and~$\ten{B}$ were derived by Goedbloed~(\cite{Go-1975},~\cite{Go-1997}) for static equilibria.

The matrices~$\ten{R}$ and~$\ten{C}$ enter if there is toroidal flow and/or external gravity. The expression for 
these matrices are 
{\renewcommand{\arraystretch}{2}
\begin{equation}
  \ten{R} =
  \begin{pmatrix}
    -\rho\left\{ \left[ J\dD\left(\dfrac{1}{J}\lambda\right)\right] - \dfrac{R\kt}{\Btheta}\left[\D\Omega^{2}\right] \right\}  &
      \left( \im\D I\mu + \lambda\G \right) \dfrac{\rho}{B^{2}}                                                                &
      \left( \im\D  \mu + \lambda\F \right) \rho                                                                               \\
    \dfrac{\rho}{B^{2}}\left( \im I\mu\dfrac{1}{J}\Dd J + \G\lambda \right)                                                    &
      -\dfrac{\rho I^{2}}{B^{4}}\left[ \dfrac{1}{J}\ptheta\mu \right]                                                          &
      -\rho\left( \dfrac{I}{B^{2}}\left[ \dfrac{1}{J}\ptheta\mu \right] - \im n \mu \right)                                    \\
    \rho\left( \im\mu\dfrac{1}{J}\Dd J + \F\lambda \right)                                                                     &
      -\rho\left( \dfrac{I}{B^{2}}\left[ \dfrac{1}{J}\ptheta\mu \right] + \im n \mu \right)                                    &
      -\rho\left[ \dfrac{1}{J}\ptheta\mu \right]
  \end{pmatrix},
  \label{eq:matrixR}
\end{equation}
}
{\renewcommand{\arraystretch}{2}
\begin{equation}
  \ten{C} =
  \begin{pmatrix}
    0                                                   &
      -\dfrac{R\kt}{\Btheta}\dfrac{\Btheta^{2}}{B^{2}}  &
       \dfrac{R\kt}{\Btheta}\dfrac{I}{R^{2}}            \\
    -\dfrac{R\kt}{\Btheta}\dfrac{\Btheta^{2}}{B^{2}}    &
      0                                                 &
      \im \left[ \dfrac{R}{J}\ptheta R \right]          \\
    \dfrac{R\kt}{\Btheta}\dfrac{I}{R^{2}}               &
      -\im \left[ \dfrac{R}{J}\ptheta R \right]         &
      0
  \end{pmatrix},
  \label{eq:matrixC}
\end{equation}
}
where
\begin{align}
  \label{eq:pressuretheta}
  \mu     & \equiv \left[ \frac{R}{J}\ptheta R \right] \Omega^{2}  - \frac{1}{J}\ptheta \Phi = \frac{1}{\rho J}\ptheta p  = \frac{1}{\rho}\vf{\Btheta}\cdot\grad{p}, \\
  \lambda & \equiv \frac{R\kt}{\Btheta}\Omega^{2} - \D\Phi.
\end{align}
Notice that the term~$\mu$ represents the pressure variation on a flux surface. The matrix~$\ten{C}$ represents the 
Coriolis effect due to the rotation while the matrix~$\ten{R}$ contains rotational as well as gravitational effects. The 
case without an external gravitational potential has been discussed by van der Holst et al.~(\cite{HBG}).

Two kinds of cylindrical limits can be obtained from the spectral equation~\eqref{eq:spectraleqn}. The first one is 
the infinitely slender torus, meaning that the radial position of accretion disk is taken to be at infinity. In that 
case the matrices $\ten{R}$ and $\ten{C}$ disappear. This is due to the fact that all equilibrium quantities will 
be become independent of the angle~$\vartheta$, the gravitational potential at infinity is zero, and the toroidal 
curvature~$\kt$ becomes zero. The flow enters only as a Doppler shift,~$-n\Omega(\psi)$, in the Doppler shifted 
eigenfrequency~$\omt(\psi)$.

The other limit is the cylindrical limit of a \emph{thin} ($|Z|/R \ll 1$) slice of plasma at the equatorial plane of 
the accretion disk. In this region all equilibrium quantities only depend on the radius~$R$, approximately. The 
matrices~$\ten{R}$ and~$\ten{C}$ do not disappear as in the previous limit. Also the toroidal curvature~$\kt$ is 
non--zero but instead the poloidal curvature~$\kp = 0$. In this case the spectral equation~\eqref{eq:spectraleqn} 
reduces to the matrix equation for a one-dimensional accretion disk presented by Keppens et al.~(\cite{KCG}).

\section{Continuous MHD spectrum \label{sec:continuous}}
In the previous section, the spectral equation~\eqref{eq:spectraleqn} governing all MHD waves and instabilities in
axisymmetric  accretion disks has been derived. This equation is the starting point to all following MHD spectral 
computations. In particular, we can derive the equations for the continuous MHD spectrum by considering modes 
localised on a particular flux surface~$\psi = \psi_{0}$. Van der Holst et al.~(\cite{HBG-L}) showed that the 
toroidal flow can drive the continuous spectrum unstable or overstable for non--gravitating plasma tori. Here, we 
show that the combination of toroidal flow and gravity  can also drive the MHD continua overstable.

\subsection{General formalism}
To derive the equations for the continuous spectrum, the normal derivative~$(\partial / \partial \psi)$ of the 
eigenfunctions is considered to be large compared to the eigenfunctions themselves and the Doppler shifted 
eigenfrequency~$\omt (\psi)$ is assumed finite. In this case, the first row of the spectral 
equation~\eqref{eq:spectraleqn} can be solved approximately,
\begin{equation}
  \frac{1}{J}\Dd JX \approx \frac{-\gamma p}{\gamma p + B^{2}} \left( \G \frac{1}{B^{2}} Y + \F Z \right)
                            - \frac{1}{\gamma p + B^{2}}\G Y
			    + \frac{\im \rho\mu}{\gamma p + B^{2}} \left( \frac{I}{B^{2}}Y + Z \right).
  \label{eq:approxX}
\end{equation}
Notice, that this solution implies that $\partial X / \partial \psi$, $Y$, and $Z$ are of the same order, but more
importantly that $X$ is small compared to $Y$ and $Z$. This means that the continuum modes are mainly tangential to a
particular flux surface.

Inserting the expression~\eqref{eq:approxX} into the second and third row of the spectral equation~\eqref{eq:spectraleqn}
results in an eigenvalue problem which is independent of the normal derivative. Hence, the reduced problem becomes
non--singular. Exploiting this property, we can write the projected displacement field components~$Y$ and $Z$ as follows:
\begin{equation}
  \begin{aligned}
    Y & \approx \delta (\psi - \psi_{0}) \eta (\vartheta), \\
    Z & \approx \delta (\psi - \psi_{0}) \zeta(\vartheta),
  \end{aligned}
\end{equation}
where $\delta (\psi - \psi_{0})$ is the Dirac delta function. Here, $Y$ and $Z$ have been split in an improper 
($\psi$--dependence) and proper part ($\vartheta$--dependence). This kind of splitting has been introduced by 
Goedbloed~(\cite{Go-1975}) for static equilibria without gravity. The reduced, non--singular eigenvalue problem 
becomes
\begin{equation}
  \left( \ten{a} + \ten{r} + 2\rho\omt\Omega\ten{c} - \rho\omt^{2}\ten{b} \right)
  \begin{pmatrix} \eta \\ \zeta \end{pmatrix} = 0,
  \label{eq:continuousspectrum}
\end{equation}
where
{\renewcommand{\arraystretch}{2}
\begin{equation}
  \ten{a} = 
  \begin{pmatrix}
    \F\dfrac{R^{2}\Btheta^{2}}{B^{2}}\F + 4\dfrac{\gamma p}{\gamma p + B^{2}} \left( R\Btheta\kg \right)^{2}  &
      -2\im\dfrac{\gamma pB}{\gamma p + B^{2}} \left( R\Btheta\kg \right)\F                                   \\
    2\im\F\dfrac{\gamma pB}{\gamma p + B^{2}}\left( R\Btheta\kg \right)                                       &
      \F\dfrac{\gamma pB^{2}}{\gamma p + B^{2}}\F
  \end{pmatrix},
  \label{eq:matrixa}
\end{equation}
}
{\renewcommand{\arraystretch}{2}
\begin{equation}
  \ten{r} =
  \begin{pmatrix}
    -4\dfrac{I}{B^{2}}\dfrac{\rho\mu B}{\gamma p + B^{2}} \left( R\Btheta\kg \right) + \dfrac{\rho I^{2}}{B^{2}}N_{\mathrm{m,pol}}^{2}  &
      \im I\dfrac{\rho\mu B}{\gamma p + B^{2}} B^{2}\F\dfrac{1}{B^{2}} + \rho I N_{\mathrm{m,pol}}^{2}                                  \\
    -\im\dfrac{I}{B^{2}}\F\dfrac{\rho\mu}{\gamma p + B^{2}} B^{2} + \rho I N_{\mathrm{m,pol}}^{2}                                       &
      -B^{2}\left[ \dfrac{1}{JB^{2}}\ptheta\dfrac{\rho\mu}{\gamma p + B^{2}} B^{2} \right] + \rho B^{2} N_{\mathrm{m,pol}}^{2}
  \end{pmatrix},
  \label{eq:matrixr}
\end{equation}
}
{\renewcommand{\arraystretch}{2}
\begin{equation}
  \ten{c} =
  \begin{pmatrix}
    0                                         &
      i \left[ \dfrac{R}{J}\ptheta R \right]  \\
    -i \left[ \dfrac{R}{J}\ptheta R \right]   &
      0
  \end{pmatrix},
  \label{eq:matrixc}
\end{equation}
}
{\renewcommand{\arraystretch}{2}
\begin{equation}
  \ten{b} =
  \begin{pmatrix}
    \dfrac{R^{2}\Btheta^{2}}{B^{2}}  &
      0                             \\
    0                               &
      B^{2}
  \end{pmatrix}.
  \label{eq:matrixb}
\end{equation}
}
In these expressions~$\kg$ is the geodesic curvature,
\begin{equation}
    \kg = \vfg{\kappa}\cdot\vfg{\pi}          
        = \frac{I}{JR\Btheta B^{2}}\ptheta B.
\end{equation}
Here, $\vfg{\kappa} = \vf{b}\cdot\grad{\vf{b}}$ is the field line curvature and $N_{\mathrm{m,pol}}^{2}$ is the 
magnetically modified Brunt-V\"ais\"al\"a frequency projected on a flux surface,
\begin{equation}
  \begin{aligned}
    N_{\mathrm{m,pol}}^{2} & = \frac{\mu}{B^{2}}\left[ \frac{1}{J\rho}\ptheta\rho - \frac{\rho\mu}{\gamma p + B^{2}} \right] \\
                           & = \left[ \frac{\vf{\Btheta}\cdot\grad{p}}{\rho B} \right] 
                               \left[ \frac{\vf{\Btheta}}{\rho B}\cdot\left( \grad{\rho}  - \frac{\rho}{\gamma p + B^{2}} \grad{p} \right) \right].
  \end{aligned}
\end{equation}
This magnetically modified Brunt-V\"ais\"al\"a frequency is similar as the one presented by van der
Holst et al.~(\cite{HBG}) for tokamaks with purely toroidal flow and as the one for static gravitational plasmas
published by~Poedts et al.~(\cite{PHG}) and Beli\"en et al.~(\cite{BPG}). Furthermore, notice that the poloidal
variation of the pressure presented by the matrix~$\ten{r}$ shows up in all its matrix elements. This destroys the
diagonal dominance of the matrix~$\ten{a}$, which corresponds to the separation of Alfv\'en and slow continuum modes.
Similar as the matrix~$\ten{C}$ in spectral equation~\eqref{eq:spectraleqn}, the matrix~$\ten{c}$ represents the Coriolis
effect.

The non--singular eigenvalue problem~\eqref{eq:continuousspectrum} is solved for poloidally periodic boundary conditions
for the eigenfunctions~$\eta$, $\zeta$. Solving this problem for a given toroidal mode number~$n$ on each flux 
surface separately results in a set of discrete eigenvalues. All these discrete sets together map out the continuous 
spectra. In the MHD formulation with the primitive variables~$\rho$, $\vf{v}$, $p$, $\vf{B}$, an additional entropy 
continuum~$n\Omega$ is found. This Eulerian entropy continuum does not couple to any of the other  continua or stable 
and unstable modes (Goedbloed et al.~\cite{GBHK-entropy}).

\subsection{Spectral properties and stability criterion}
In this subsection a stability criterium for the continua will be derived. For this derivation we need to construct a
Hilbert space with appropriate inner product. The parallel gradient operator~$\F$ is a Hermitian operator under the
inner product
\begin{equation}
  \oint J v^{*} \left( \F w \right) \dd{}\vartheta = \oint J \left( \F v \right)^{*} w \dd{}\vartheta
  \label{eq:HermitianF}
\end{equation}
for poloidally periodic functions $v$ and $w$. Inspired by this property a Hilbert space can be defined with the
inner product
\begin{equation}
  \langle \vf{v},\vf{w} \rangle \equiv \oint \vf{v}^{*}\cdot\vf{w} J \dd{}\vartheta .
  \label{eq:innerproduct}
\end{equation}
In this Hilbert space, the matrix operators~$\ten{a}$, $\ten{b}$, $\ten{c}$, and~$\ten{r}$ are Hermitian operators. 
By setting~$\vf{v}$ and~$\vf{w}$ equal to the eigenfunctions~$(\eta,\zeta)^{\mathrm{T}}$, one can derive a quadratic 
polynomial for the eigenfrequency~$\omt$ from the spectral equation~\eqref{eq:continuousspectrum} for the continua.
\begin{equation}
  a\omt^{2} - 2b\omt - c = 0.
  \label{eq:polynomial}
\end{equation}
In this equation, the coefficients are
\begin{align}
  \label{eq:ointa}
  a  & \equiv \oint \rho\left( \frac{R^{2}\Btheta^{2}}{B^{2}}|\eta|^{2} + B^{2}|\zeta|^{2} \right)               J\dd{}\vartheta,  \\
  \label{eq:ointb}
  2b & \equiv 2i\oint \rho\Omega\left[ \frac{R}{J}\ptheta R \right] \left( \eta^{*}\zeta - \eta\zeta^{*} \right) J\dd{}\vartheta,  \\
  \label{eq:ointc}
  c  & \equiv \oint \left\{   \frac{R^{2}\Btheta^{2}}{B^{2}} \left| \F\eta \right|^{2}
                            + \frac{\gamma pB^{2}}{\gamma p + B^{2}} \left| \F\zeta + 2\im\frac{R\Btheta\kg}{B}\eta - \im\frac{\rho\mu}{\gamma p}\left(\frac{I}{B^{2}}\eta + \zeta \right) \right|^{2} \right. \\
     & \phantom{\equiv \oint \left\{\right.} \left.
	   		    + \rho B^{2}N_{\mathrm{BV,pol}}^{2} \left| \frac{I}{B^{2}}\eta + \zeta \right|^{2} \right\} J\dd{}\vartheta, \nonumber
\end{align}
where $N_{\mathrm{BV,pol}}^{2}$ is the Brunt-V\"ais\"al\"a frequency projected on a flux surface,
\begin{equation}
  \begin{aligned}
    N_{\mathrm{BV,pol}}^{2} & = \frac{\mu}{B^{2}}\left[ \frac{1}{J\rho}\ptheta\rho - \frac{\rho\mu}{\gamma p} \right]  \\
                         & =  \left[ \frac{\vf{\Btheta}\cdot\grad{p}}{\rho B} \right] 
                              \left[ \frac{\vf{\Btheta}}{\rho B}\cdot\left( \grad{\rho}  - \frac{\rho}{\gamma p} \grad{p} \right) \right] \\
			 & = -\left[ \frac{\vf{\Btheta}\cdot\grad{p}}{\rho B}    \right]
                              \left[ \frac{\vf{\Btheta}\cdot\grad{S}}{\gamma BS} \right].
  \end{aligned}
\end{equation}
Note that the coefficient~$a$ is always non zero but, more importantly, that the coefficients~$a$, $b$, $c$ are real 
due to the Hermitian property of the inner product. Solutions of this polynomial 
are~$\omt = (b \pm \sqrt{b^{2} + ac}) / a$. This means that if $b^{2} + ac \ge 0$, the solutions are waves with real
frequencies. But if $b^{2} + ac < 0$, the solutions contain a damped stable wave and an overstable mode. For the 
complex solution~$\omt$ one derives its absolute value~$|\omt|=\sqrt{-c / a}$ by taking the linear combination of the 
polynomial with its conjugate. Furthermore, $\mathrm{Re}(\omega)=n\Omega + b/a$. It can easily be shown that the 
continuum is stable if~$c \ge 0$. The coefficient~$c$ is always equal or greater than zero if
\begin{equation}
  N_{\mathrm{BV,pol}}^{2} \ge 0
  \label{eq:stabilitycriterion}
\end{equation}
is satisfied everywhere in the plasma. From this stability criterion, we conclude that equilibria with $S = S(\psi)$ 
and equilibria with $T = T(\psi)$, $\gamma \ge 1$ are always stable. Equilibria where the density
is a flux function violate this criterion, which may result in the appearance of damped stable and overstable modes.
Notice that the stability criterion~\eqref{eq:stabilitycriterion} is analogous to the Schwarzschild criterion for
convective instability. The only important difference is that in the Schwarzschild criterion one deals with normal 
derivatives while this criterion deals with tangential ones. This has also been noticed by Hellsten \& 
Spies~(\cite{HeSp}), Hameiri~(\cite{Ham}), and Poedts et al.~(\cite{PHG}).

\section{Numerical codes \label{sec:codes}}
For the stability analysis of axisymmetric accretion tori two numerical codes, the equilibrium code
FINESSE (Beli\"en et al.~\cite{BBGHK}) and the spectral code PHOENIX~(Blokland et al.~\cite{BHKG}), have been used. 
In the next two subsections their algorithmic details will be briefly discussed.

\subsection{The equilibrium code FINESSE}
\begin{figure}
  \centering
  \includegraphics[trim=50 75 0 100, width=0.7\textwidth]{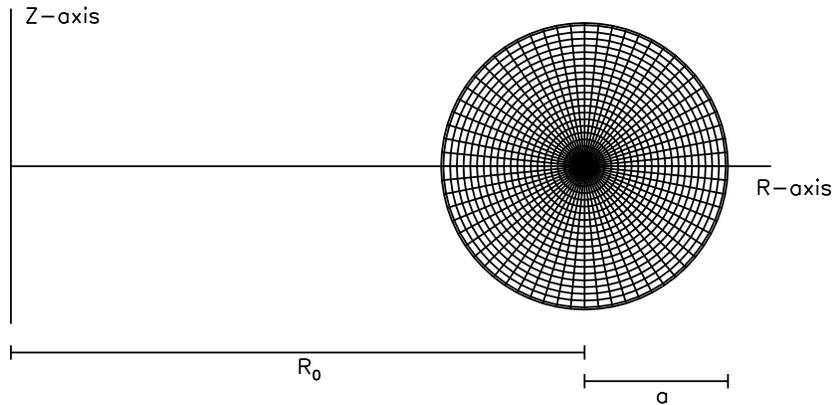}
  \caption{An example of the geometry of an accretion disk with a circular cross-section. Here,~$R_{0}$ and~$a$ are the
           major radius of the geometric axis and the radius of the last closed flux surface, respectively}
  \label{fig:geometry}
\end{figure}

The FINESSE code developed by Beli\"en et al.~(\cite{BBGHK}) can compute a stationary axisymmetric, gravitating MHD
equilibrium described by the extended Grad--Shafranov equations~\eqref{eq:gradshafranovT}, \eqref{eq:gradshafranovrho},
or~\eqref{eq:gradshafranovS} for a given poloidal cross--section and inverse aspect ratio~$\epsilon \equiv a / R_{0}$.
Here,~$a$ and~$R_{0}$ are the minor radius of the last closed flux surface and the major radius of the geometric axis,
respectively. Fig.~\ref{fig:geometry} shows an example of the geometry of an accretion disk with a circular
cross--section as seen in a poloidal plane with the central gravitational object at the origin. The extended
Grad--Shafranov equation has been discretized using a finite element method in combination with the standard 
Galerkin method. The elements used are isoparametric bicubic Hermite ones. These bicubic elements ensure that the 
computed solution has the desired high accuracy (fourth order) needed for the stability analysis. FINESSE solves the
elliptic PDE problem using the Picard iteration scheme. The imposed boundary conditions assume that the fixed 
boundary represents the last closed flux surface. The same FINESSE code can actually handle the more general case
of stationary MHD equilibrium with non--vanishing poloidal flows as well, but here we restrict our discussion to 
equilibria with purely toroidal flows.

\subsection{The spectral code PHOENIX}
The spectral code PHOENIX developed by van der Holst (Blokland et al.~\cite{BHKG}) is used for the stability analysis 
of a two dimensional accretion disk. PHOENIX can compute the complete MHD spectrum for a given two dimensional 
equilibrium, toroidal mode number~$n$, and a range of poloidal mode numbers~$m$. A range of poloidal mode numbers is 
needed due to the mode coupling caused by allowing for a non--circular cross--section, gravitational stratification, 
and the toroidal geometry of the accretion disk. PHOENIX does not directly solve the spectral 
equation~\eqref{eq:spectraleqn}, but instead  employs the linearised version of the full set of MHD 
Eqs.~\eqref{eq:momentum}--\eqref{eq:mass}. These linearised equations are then discretized using a finite element 
method in the normal $\psi$--direction, a spectral method in the poloidal direction and a standard Galerkin method is 
used to obtain a linear generalised eigenvalue problem for the eigenfrequencies~$\omega$. The elements used in the 
$\psi$--direction for the perturbed quantities are a combination of quadratic and cubic Hermite elements to prevent 
the creation of spurious eigenvalues (Rappaz~\cite{Ra}). This results in a non--Hermitian eigenvalue problem, which is 
solved using the Jacobi--Davidson method (Sleijpen \& van der Vorst~\cite{SH}). The boundary conditions used for the 
perturbations are those of a perfect conducting wall. Strictly speaking, these conditions can not be applied to waves 
and instabilities that appear inside an accretion disk. However, if the wave phenomena are sufficiently localised or 
not significantly affected by the position of the wall, these boundary conditions have hardly any influence. This is 
particularly true for the continuous branches of the MHD spectrum on which we concentrate in what follows. These modes 
are purely localised on a flux surface, and are not affected by the wall. Like FINESSE, PHOENIX can handle poloidal
flows as well, but we have exploit it here for equilibria with toroidal flow only.

The continuous spectrum is computed by a less expensive method introduced by Poedts \& Schwarz~(\cite{PS}). In this method,
on each individual flux surface the cubic Hermite and the quadratic elements are replaced by $\log (\epsilon)$
and~$1/\epsilon$, respectively. Due to this replacement, the singular behaviour of the continuous spectrum has been
approximated by the perturbed quantities. This singular behaviour has been described by Goedbloed~(\cite{Go-1975}) 
and Pao~(\cite{Pa}) for static, axisymmetric and toroidal plasma. Recently, the analogous treatment has been used
by Goedbloed et al.~(\cite{GBHK}) for plasmas with toroidal and poloidal flows and gravity. The resulting eigenvalue
problem is then a small algebraic problem of a size set by the range in poloidal mode numbers~$m$ on each flux surface,
and is then solved using a direct QR method and a scan over all flux surfaces. This yields detailed information on all
MHD continua.

\section{Accretion disks with density as a flux function \label{sec:results}}
We present our numerical results on disk equilibria and stability using the codes described in the previous section.
For the equilibrium computations the ratio of the specific heats $\gamma = 5/3$ and the gravitational potential is an
external Newtonian one, i.e.~$\Phi = -GM_{*} / \sqrt{R^{2} + Z^{2}}$, where $G$ and $M_{*}$ are the gravitational 
constant and the mass of the central object, respectively. Furthermore, we assume  that the density is a flux function
for all equilibrium computations. From our theoretical analysis above, this is the relevant case which may give rise to 
damped stable and overstable continuum mode pairs due to the presence of toroidal flow and gravity.

\subsection{Thin accretion disk with constant density}
\begin{figure}[ht]
  \centering
  \includegraphics[width=0.6\textwidth,clip]{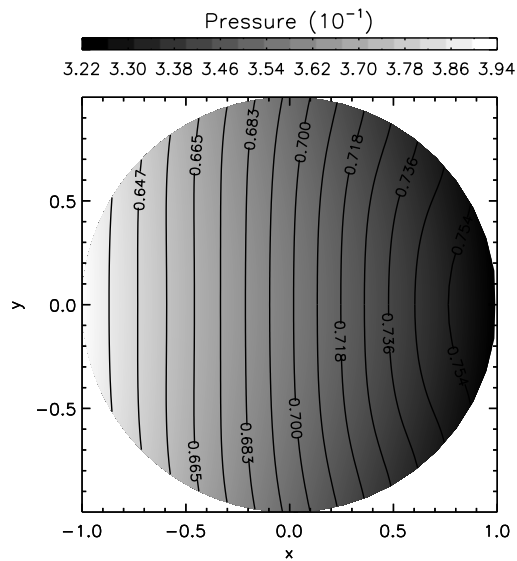}
  \caption{The two--dimensional pressure (gray--scale) and plasma beta~$\beta= 2p/B^{2}$ (contours) profile for an 
           accretion disk with constant density. The cross--section is circular and the inverse aspect
	   ratio~$\epsilon=0.1$. The central object is to the left of the figure.}
  \label{fig:rho-pressure}
\end{figure}

As the first case, we take an accretion disk with constant density as also presented by Blokland et al.~(\cite{BHKG}).
The equilibrium is described by the following flux functions:
\begin{equation}
  \begin{aligned}
    I^{2}(\psi) & = A (1 - 0.0385\psi + 0.02\psi^{2} + 0.00045\psi^{3}),  &
      \rho(\psi) & = 1,                                                   \\
    p_{0}(\psi) & = AB(1 - 0.9\psi),                                      &
      \Omega(\psi) & = C (1 - 0.9\psi^{2}),
  \end{aligned}
  \label{eq:equil-rho}
\end{equation}
where the values of the coefficients $A$, $B$, and $C$ are given by $112$, $0.01$, and $0.1$. Furthermore, the considered poloidal cross-section is circular
and the inverse aspect ratio~$\epsilon = 0.1$.  In these calculations, $GM_{*} = 1$ after scaling with respect to a
reference density, the vacuum magnetic field, and the minor radius of the accretion disk. Fig.~\ref{fig:rho-pressure}
shows the computed two--dimensional pressure and plasma beta~$\beta (= 2p/B^{2})$  profile. The pressure decreases
monotonically outwards due to the fact that the term proportional to the gravitational potential~$\Phi$ dominates in the 
pressure equation~\eqref{eq:pressurerho}. In contrast, the plasma beta increases monotonically outwards from $0.63$ at 
the inner part to $0.76$ at the outer part of the disk. The ratio~$v_{\varphi} / v_{\mathrm{Kepler}} = [ 0.085 , 1.049]$, 
where~$v_{\mathrm{Kepler}}$ is the Keplerian velocity.

\begin{figure}[ht]
  \centering
  \begin{tabular}{c}
    \includegraphics[height=0.3\textheight,clip]{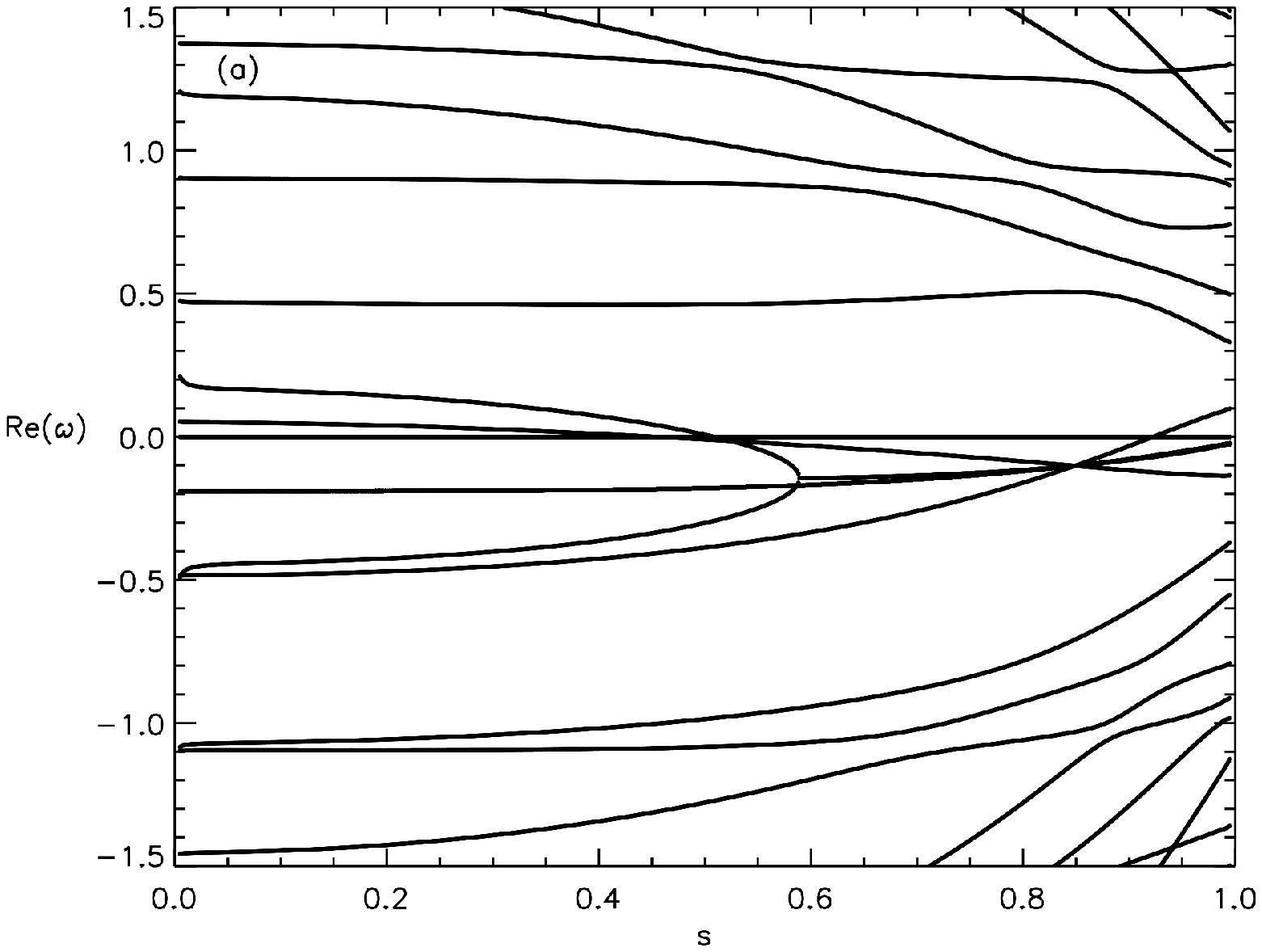}          \\
    \includegraphics[height=0.3\textheight,clip]{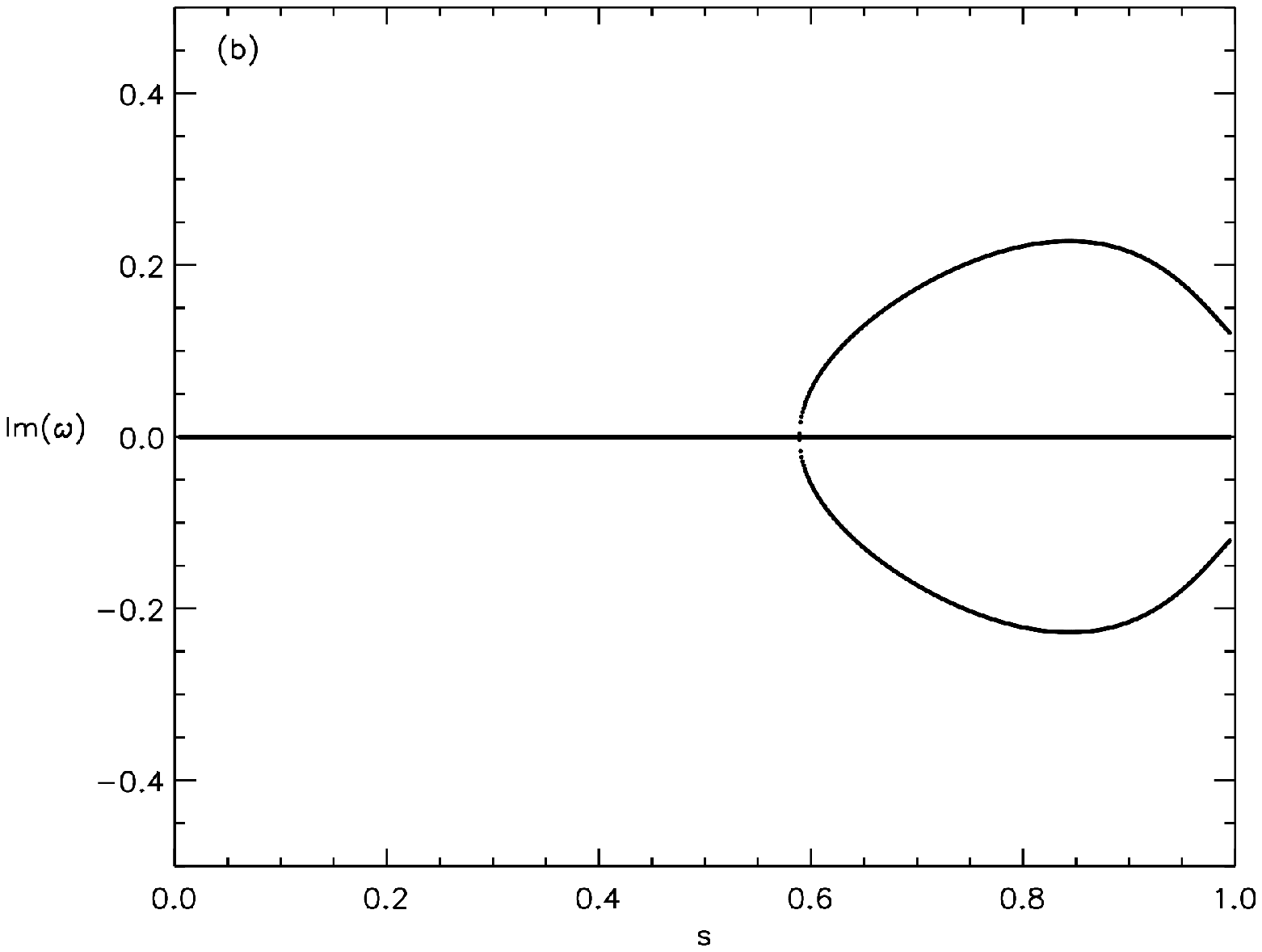}          \\
    \includegraphics[height=0.3\textheight,clip]{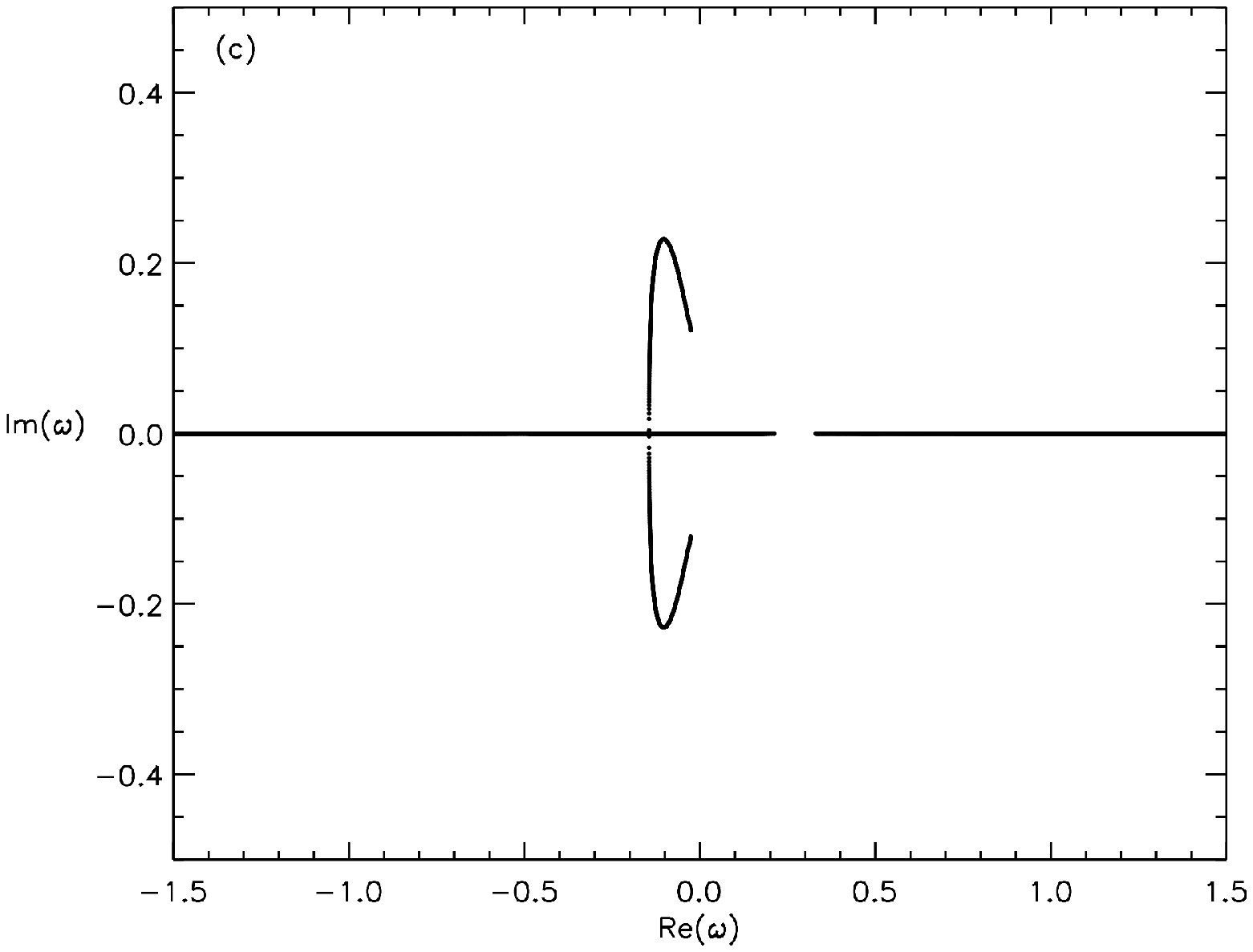} 
  \end{tabular}
  \caption{(a) Real, and (b) imaginary parts of the sub--spectrum of the MHD continua as functions of the radial
           flux coordinate~$s \equiv \sqrt{\psi}$ and (c) represented in the complex plane for toroidal mode 
	   number~$n=-1$ and poloidal mode numbers~$m=[-3,5]$. The corresponding disk equilibrium shown in
	   Fig.~\ref{fig:rho-pressure} has a constant density distribution.}
  \label{fig:rho-spectrum}
\end{figure}

The computation of the continuous spectrum of this equilibrium is done for toroidal mode number~$n=-1$ and poloidal 
mode numbers~$m=[-3,5]$. Fig.~\ref{fig:rho-spectrum} shows the sub--spectrum of the MHD continua. The plots clearly 
indicate the existence of stable and overstable modes in the MHD continua. The plots (a) and (b) show the real and 
imaginary part of the eigenfrequencies~$\omega$ that belong to the continuous spectrum as a function of the radial flux
coordinate~$s\equiv\sqrt{\psi}$. The frequencies of the continua in the complex plane are shown in the plot (c). It 
should be noted that it is difficult to label the different branches of the MHD continua by a ``dominant'' poloidal 
mode number~$m$. This could be done by making use of analytical expressions for an infinitely slender 
torus~($\epsilon \to 0$), but then the gravity of the central object would be neglected. In our case, these expressions 
can hardly be used to identify the different branches of the MHD continua, because gravity plays a dominant role in 
the equilibrium as well as in the spectral analysis. Notice that in plot (a) two branches merge at $s \approx 0.59$, 
while at the same location the imaginary part of the eigenfrequency starts to become non--zero. Intricate mode
couplings thus give rise to the emergence of pairs of overstable and damped stable modes.

\begin{figure}
  \centering
  \includegraphics[width=0.6\textwidth]{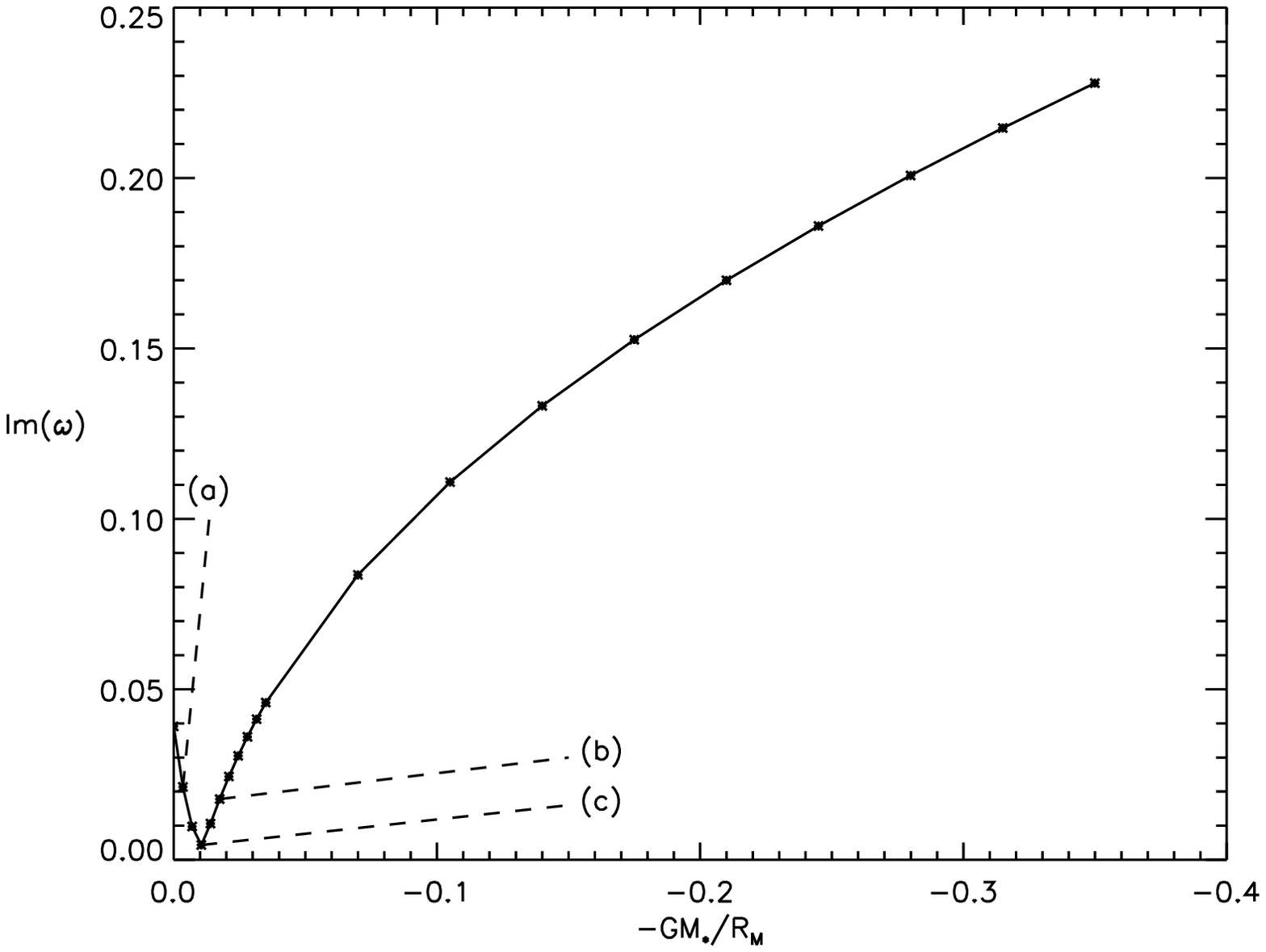}
  \caption{The growth rate of the most unstable continuum mode as a function of the value of gravitational potential on
           the magnetic axis~$R_{\mathrm{M}}$. The gravitational potential is scaled with respect to the Alfv\'en speed
	   on the magnetic axis.}
  \label{fig:growthrate-gravity}
\end{figure}

\begin{figure}[ht]
  \centering
  \begin{tabular}{ll}
    \includegraphics[width=0.6\textwidth,clip]{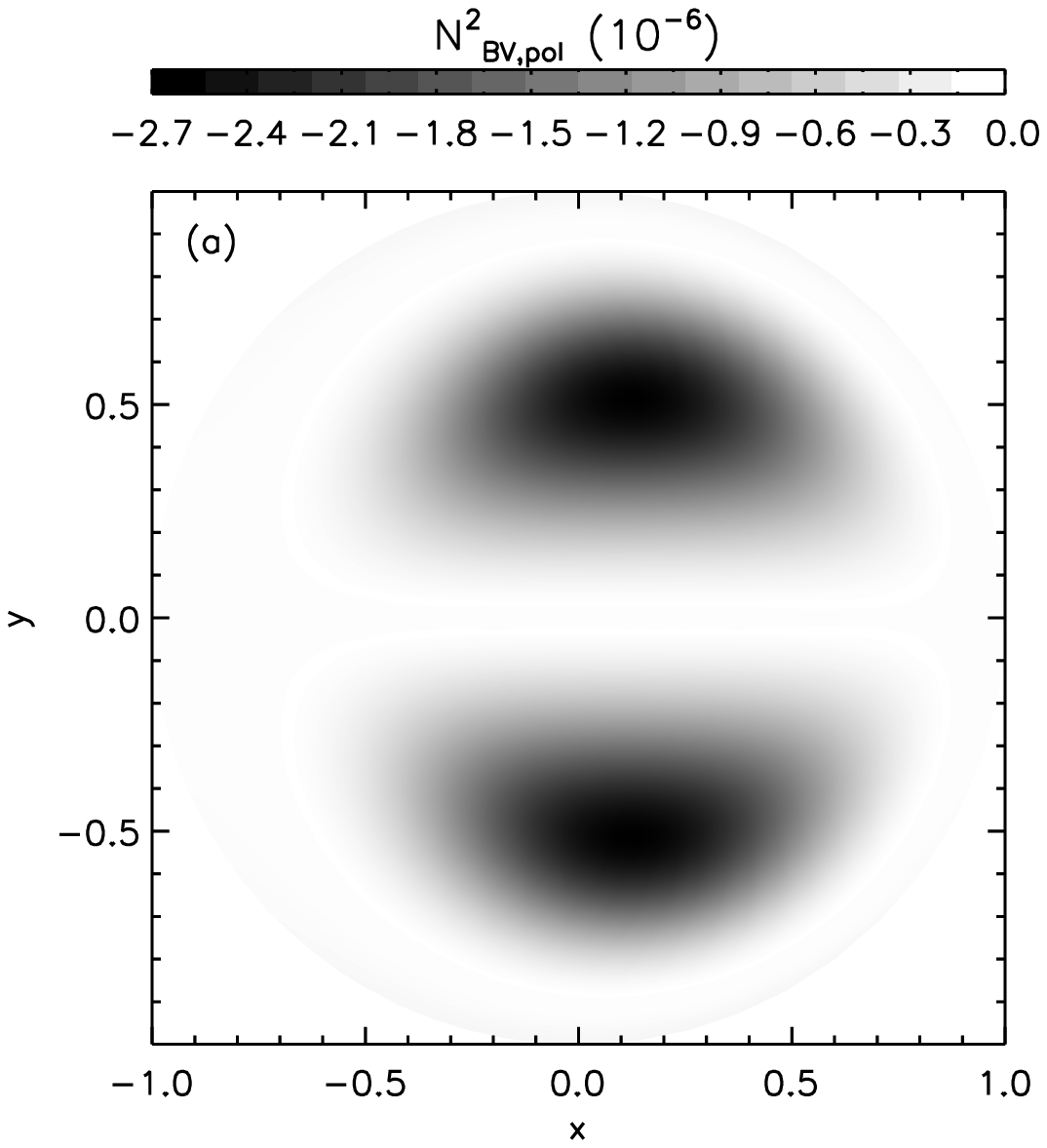}  &
    \hspace{-4cm}
    \includegraphics[width=0.6\textwidth,clip]{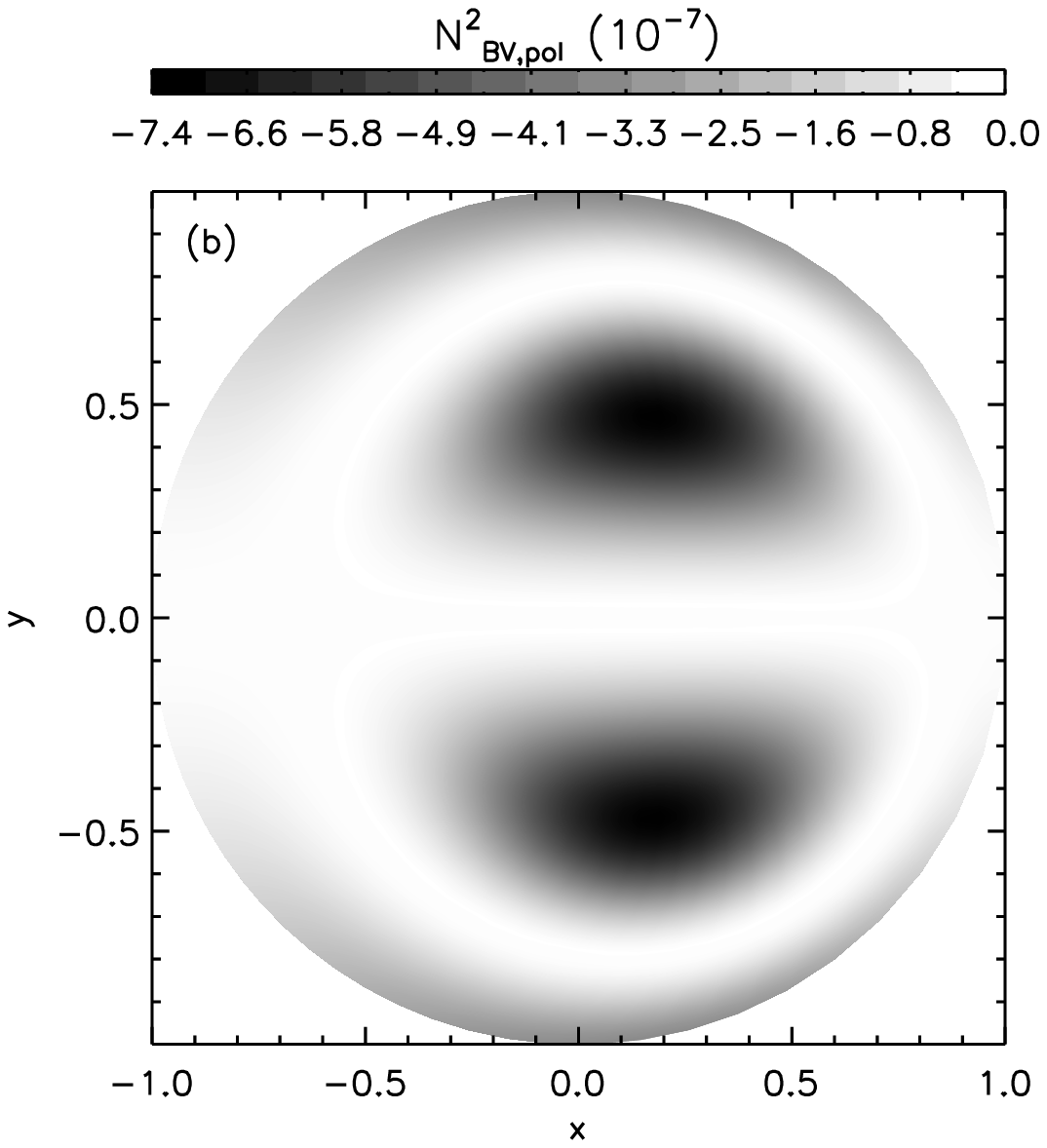}  \\
    \includegraphics[width=0.6\textwidth,clip]{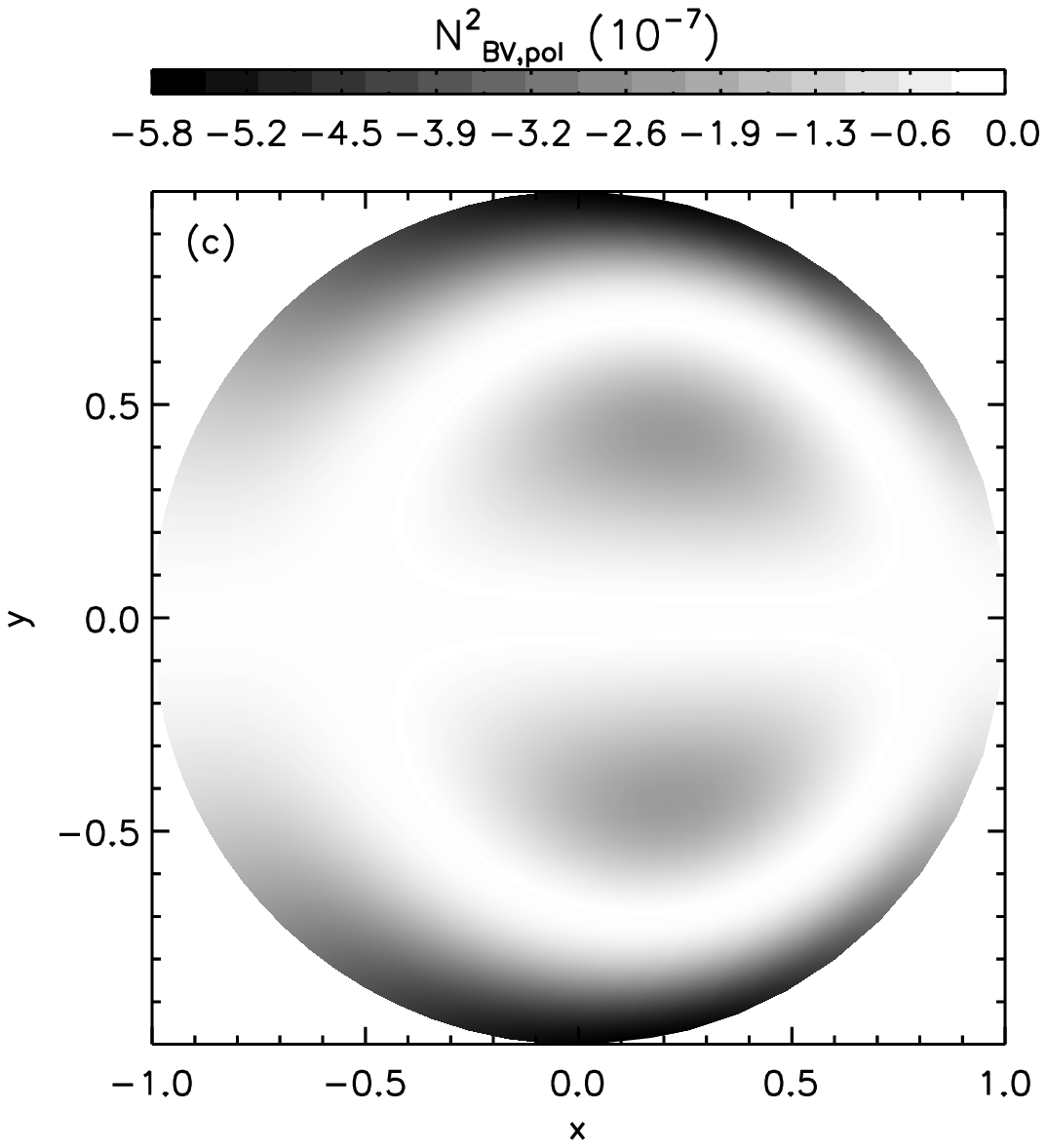}  &
  \end{tabular}
  \caption{The Brunt-V\"ais\"al\"a frequency projected on a flux surface for the points (a), (b), and (c) shown in
           Fig.~\ref{fig:growthrate-gravity} shown as a contour plot in a poloidal cross--section.}
  \label{fig:rho-N2}
\end{figure}

Next, we investigate the influence of the external gravitational potential on the overstable modes presented in 
the previous figure. This is done by varying the mass of the central object~$GM_{*}$. The result is shown in 
Fig.~\ref{fig:growthrate-gravity}, where the most overstable mode of the MHD continua is plotted as function of 
the gravitational potential on the magnetic axis~$R_{\mathrm{M}}$. Notice that each point in this plot corresponds 
to a different equilibrium calculated with FINESSE. PHOENIX is used to analyse the stability of the computed 
equilibrium. The gravitational potential is stabilizing from 0 to -0.01 and destabilizing from -0.01 to lower values. 
The zero value corresponds to a toroidally rotating, non-gravitating ``tokamak'' plasma. Its instability is in 
agreement with the results by van der Holst et al.~(\cite{HBG-L}). Fig.~\ref{fig:rho-N2} shows the projected 
Brunt-V\"ais\"al\"a frequency for the points (a), (b), and (c) indicated in Fig.~\ref{fig:growthrate-gravity}. 
Plot (a) shows that the minimum value of this frequency is reached inside the plasma, while plot (c) shows that this 
value is reached at the edge of the accretion disk. Point (b) is the transition from the minimum value inside to the 
edge of the disk as seen in plot (b).

\subsection{Thin accretion disks with non--constant density}
\begin{figure}[ht]
  \centering
  \includegraphics[width=0.6\textwidth,clip]{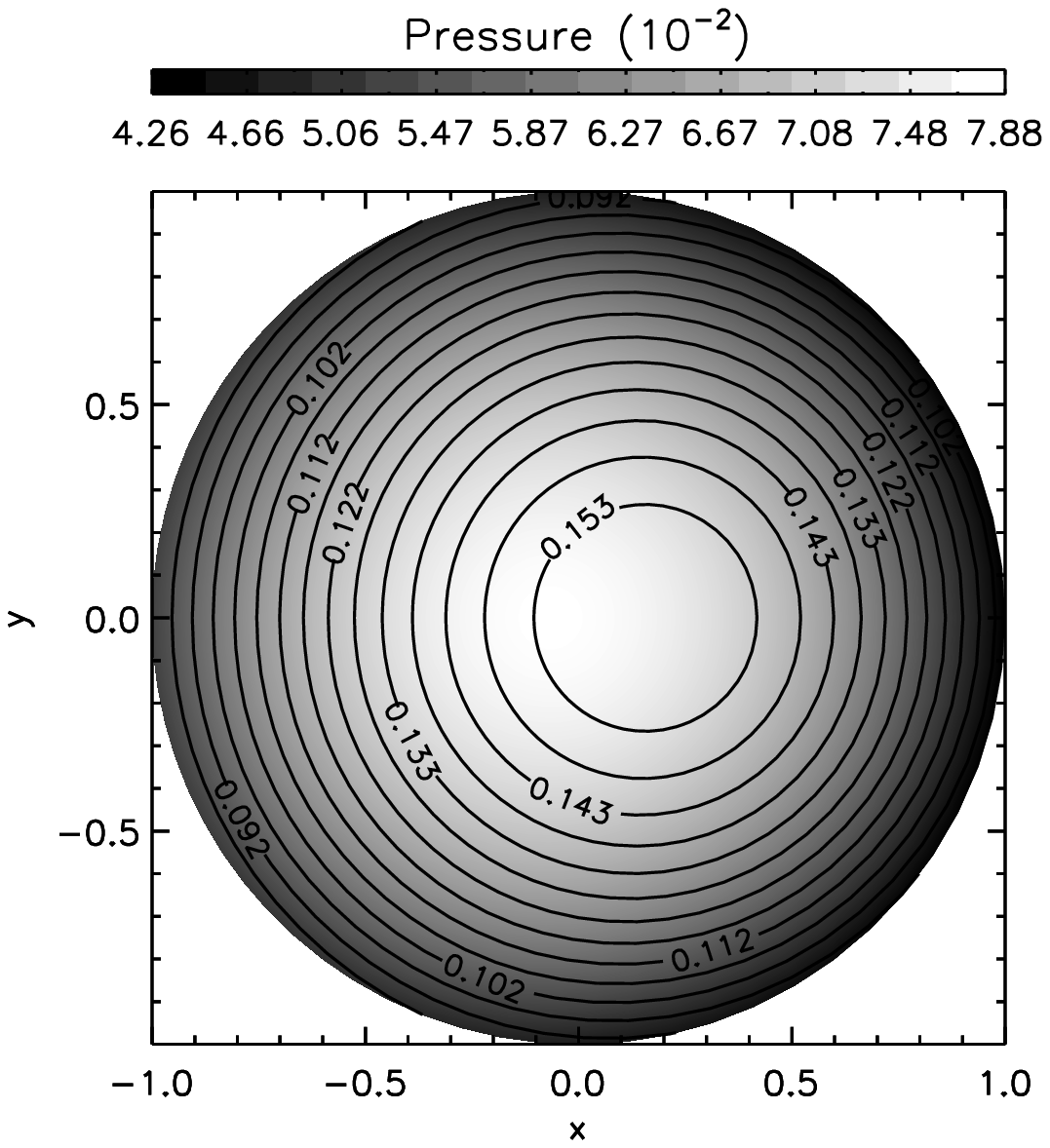}
  \caption{The pressure (gray--scale) and plasma beta~$\beta= 2p/B^{2}$ (contours) distribution in a poloidal
           cross--section for an accretion disk with density~$\rho \propto (1-0.4\psi)$ and an inverse aspect
	   ratio~$\epsilon=0.1$.}
  \label{fig:epsilon-0.10-pressure}
\end{figure}

To model a more realistic accretion disk equilibrium, we allow a density gradient across the poloidal magnetic 
surfaces. All other flux functions are kept as in the previous case, i.e.
\begin{equation}
  \begin{aligned}
    I^{2}(\psi) & = A (1 - 0.0385\psi + 0.02\psi^{2} + 0.00045\psi^{3}),  &
      \rho(\psi) & = 1 - 0.4\psi,                                         \\
    p_{0}(\psi) & = AB(1 - 0.9\psi),                                      &
      \Omega(\psi) & = C (1 - 0.9\psi^{2}),
  \end{aligned}
  \label{eq:equil-epsilon}
\end{equation}
where the coefficients $A$, $B$, $C$ are given by $91.5$, $0.01$, and $0.1$. Also in this case, the poloidal 
cross--section is circular, the inverse aspect ratio~$\epsilon=0.1$, and the scaled mass of the central object 
is~$GM_{*}=1$. The two--dimensional pressure and plasma beta profile are shown in Fig.~\ref{fig:epsilon-0.10-pressure}. 
This profile shows that the pressure reaches its maximum within the disk, instead of at the inner boundary of the disk,
as in the case of a constant density. This difference can be explained as follows. Here, also the 
term~$p_{0} \Phi(R,Z)/T_{\rho} = \rho \Phi(R,Z)$ dominates in the pressure equation~\eqref{eq:pressurerho}, but, due to 
the small inverse aspect ratio, the variation in the gravitational potential is small compared to the one in the density. 
This implies that the density, which is a flux function, mainly determines the shape of the dominant term in the
Eq.~\eqref{eq:pressurerho}. The temperature (not shown) decreases monotonically outwards. The range of plasma 
beta~$\beta=[0.082 , 0.158]$ and its maximum value is reached  within the accretion disk. These low values for
the plasma beta imply that the disk is strongly magnetized. Furthermore, the range of the 
ratio~$v_{\varphi}/v_{\mathrm{Kepler}}=[0.085 , 1.044]$.

\begin{figure}[ht]
  \centering
  \begin{tabular}{c}
    \includegraphics[height=0.3\textheight,clip]{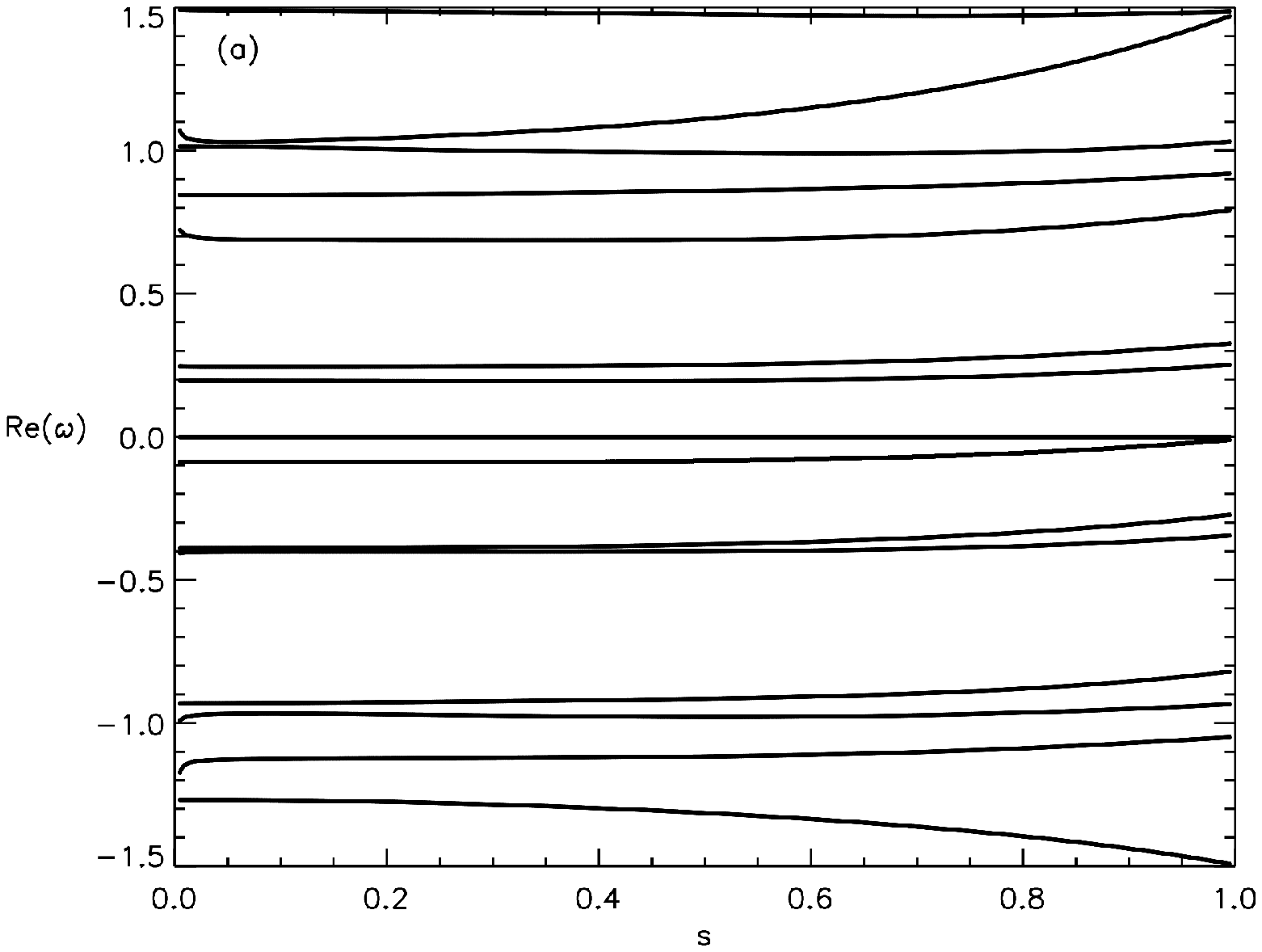}          \\
    \includegraphics[height=0.3\textheight,clip]{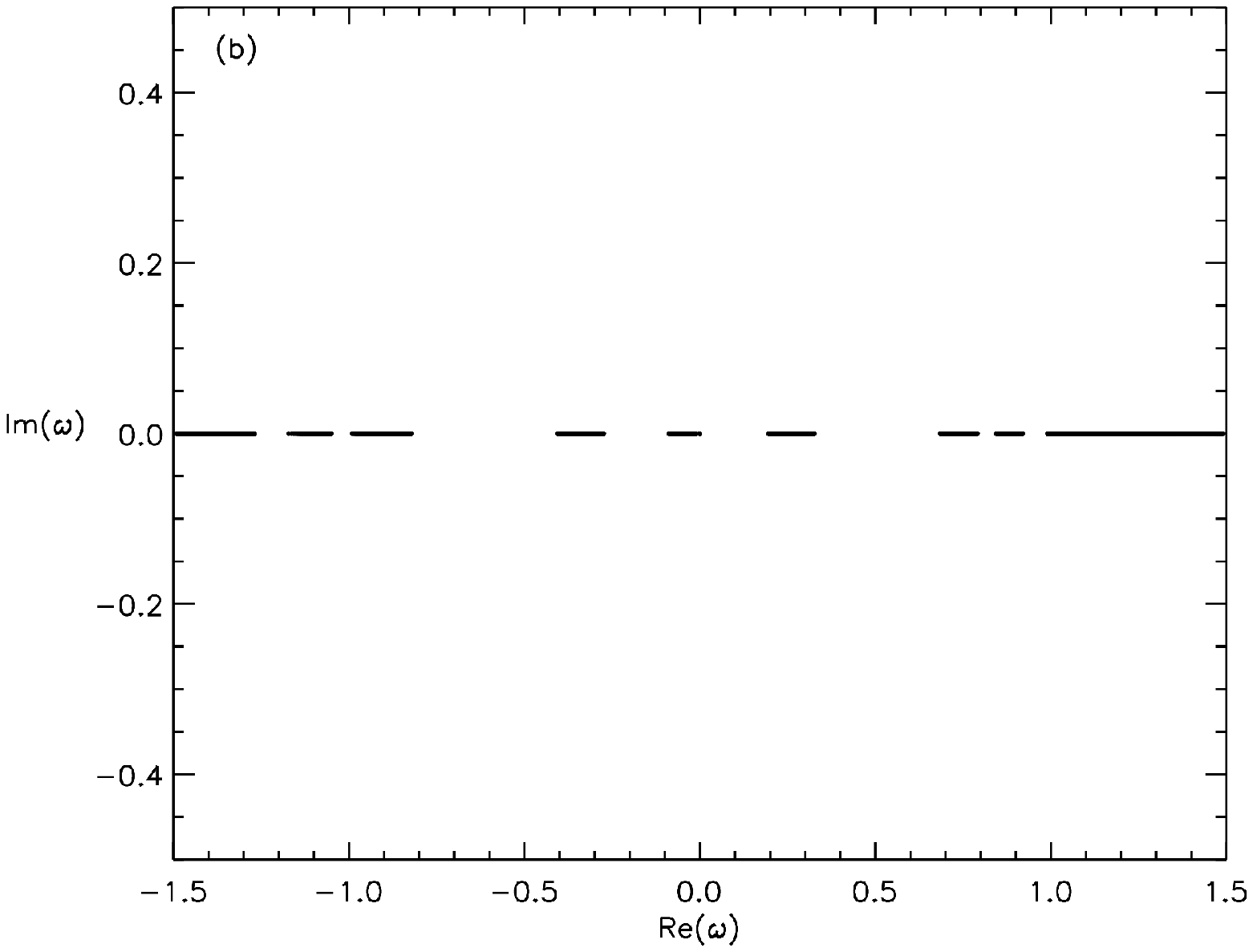}
  \end{tabular}
  \caption{For the accretion disk equilibrium shown in Fig.~\ref{fig:epsilon-0.10-pressure}, (a) the real parts of the
           sub--spectrum of MHD continua as function of the radial flux coordinate~$s \equiv \sqrt{\psi}$ and (b) 
	   sub--spectrum of the MHD continua in the complex plane are shown for toroidal mode number~$n=-1$ and 
	   poloidal mode numbers~$m=[-3,5]$. For these mode numbers, no unstable or overstable modes are found.}
  \label{fig:epsilon-0.10-spectrum}
\end{figure}

For this case, the continuous spectrum has been computed for toroidal mode number~$n=1$ and poloidal mode
numbers~$m=[-3,5]$. The results are shown in Fig.~\ref{fig:epsilon-0.10-spectrum}, which contains two plots.
In the left one, the real part of the eigenvalues is plotted against the ``radial'' coordinate~$s \equiv \sqrt{\psi}$,
while the right one shows the eigenvalues in the complex plane. For this case, there are neither purely exponential
unstable or overstable modes. Due to the presence of toroidal flow, gravity and a non--constant density, it is 
possible to create new or wider gaps between the different MHD continuum branches, with avoided crossings as a result 
of poloidal mode couplings. This has been shown by van der Holst et al.~(\cite{HBG}) for tokamak equilibria with 
toroidal flow. In these gaps, there may exist global modes similar to the Toroidal Flow--induced Alfv\'en 
Eigenmode (TFAE) for tokamaks found  by van der Holst et al.~(\cite{HBG-L}) or to the Stratification--induced 
Alfv\'en Eigenmode (SAE) for coronal flux tubes found by Beli\"en et al.~(\cite{BPG-L}). We will investigate this 
possibility for novel global modes in accretion disk context in future work.

\subsection{Thick accretion disk}
\begin{figure}[ht]
  \centering
  \includegraphics[width=0.6\textwidth,clip]{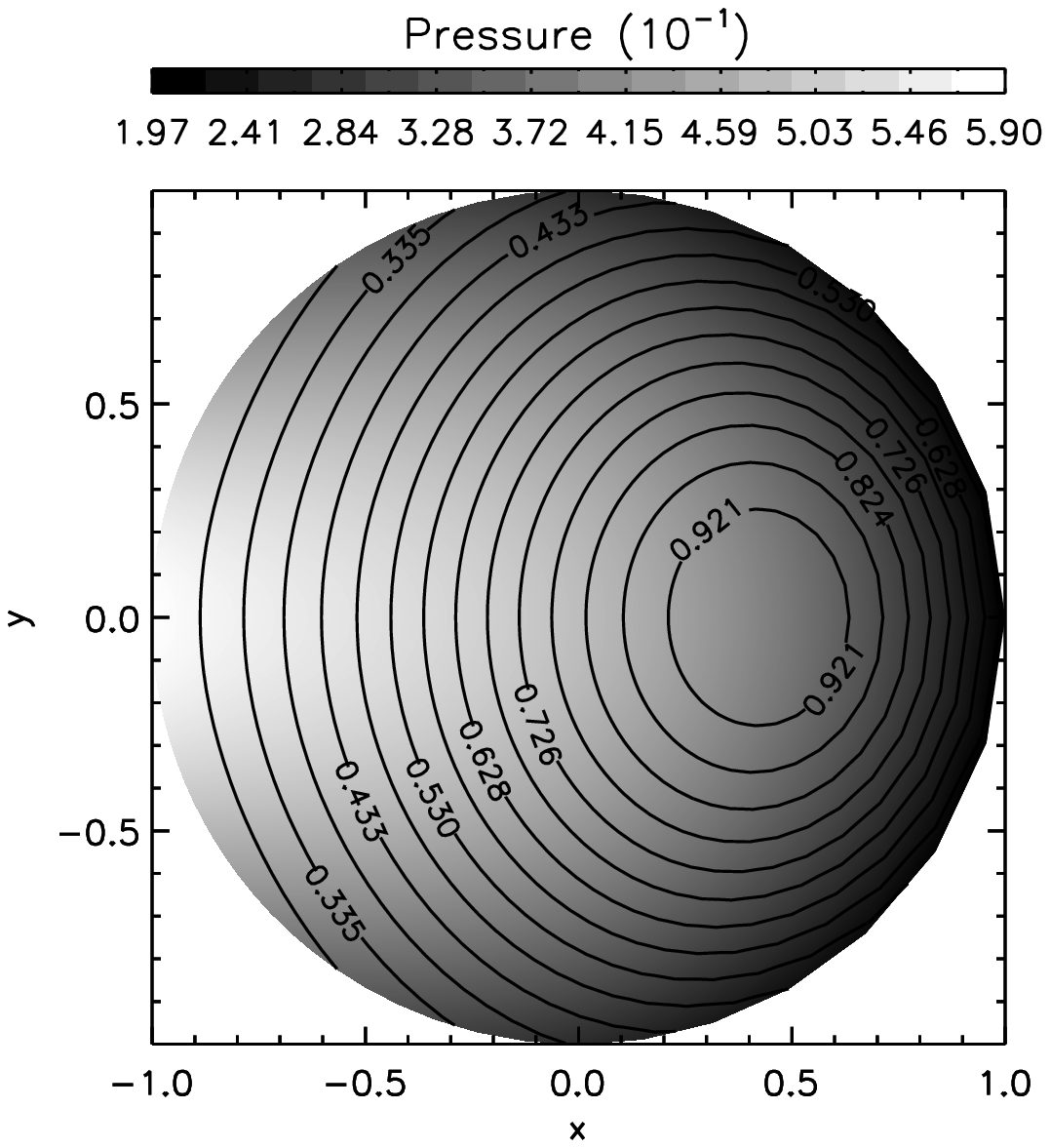}
  \caption{The pressure (gray--scale) and plasma beta~$\beta= 2p/B^{2}$ (contours) for a thick accretion disk with
           density~$\rho \propto (1-0.4\psi)$ and an inverse aspect ratio~$\epsilon=0.5$.}
  \label{fig:epsilon-0.50-pressure}
\end{figure}

Next, we increase the inverse aspect ratio~$\epsilon$ from 0.1 to 0.5, while keeping all other quantities the same. 
Again, the poloidal cross--section is chosen circular. The resulting pressure profile is presented in 
Fig.~\ref{fig:epsilon-0.50-pressure}. Here, The pressure decreases monotonically outward, similar to the equilibrium 
with constant density. Again, the term~$\rho \Phi(R,Z)$ dominates in the pressure equation~\eqref{eq:pressurerho}. 
The inverse aspect ratio is now not small so that both the gravitational potential and the density determines the 
shape of the pressure profile. Also in this case, the temperature decreases in the outward direction. The plasma 
beta~$\beta$ of this accretion disk ranges from 0.237 up to 0.970 and reaches its maximum within the disk. The 
deviation from a Keplerian disk is expressed by the ratio~$v_{\phi}/v_{\mathrm{Kepler}} = [0.007, 0.282]$. Thus the 
disk completely rotates sub-Keplerian.

\begin{figure}[ht]
  \centering
  \begin{tabular}{l}
    \includegraphics[height=0.3\textheight,clip]{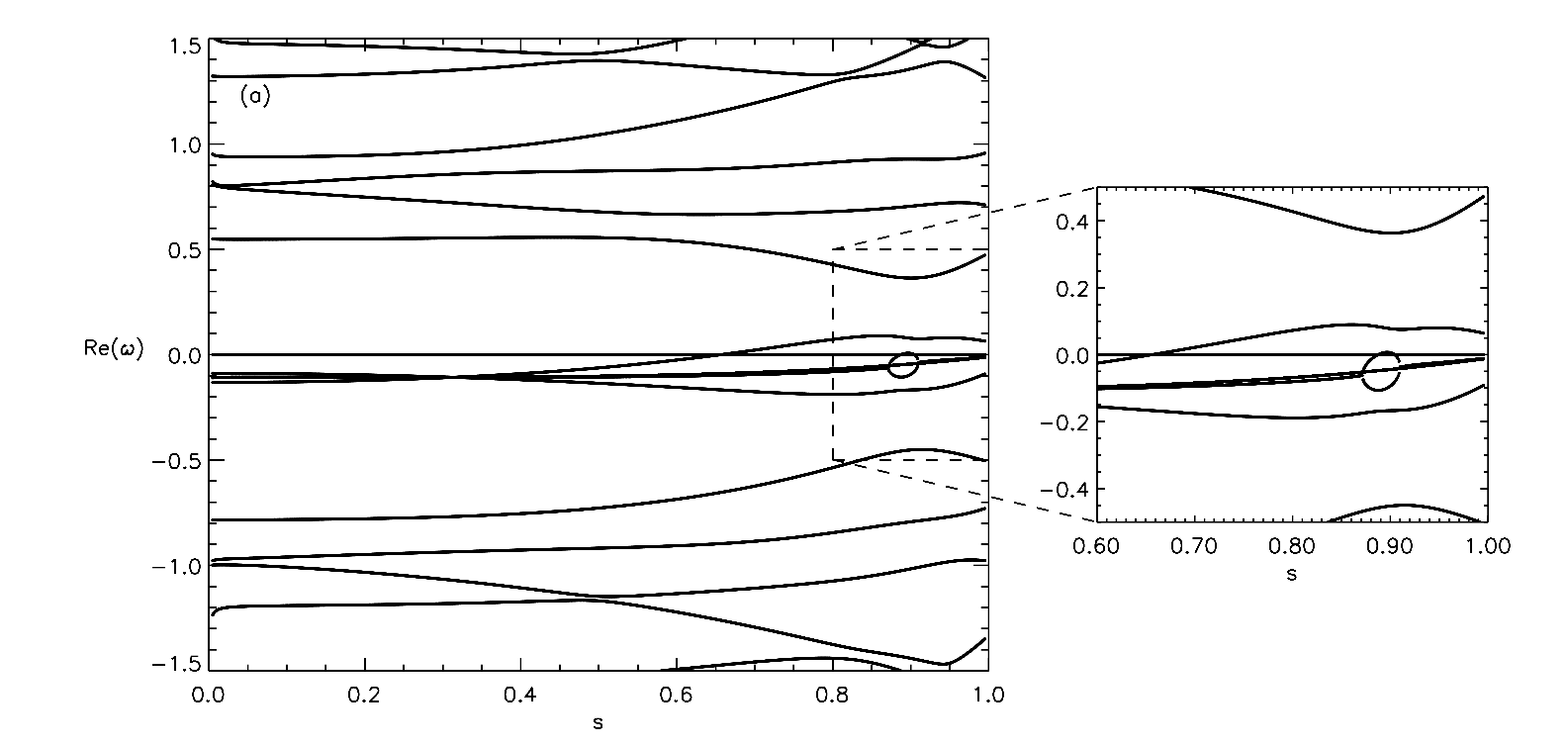}          \\
    \includegraphics[height=0.3\textheight,clip]{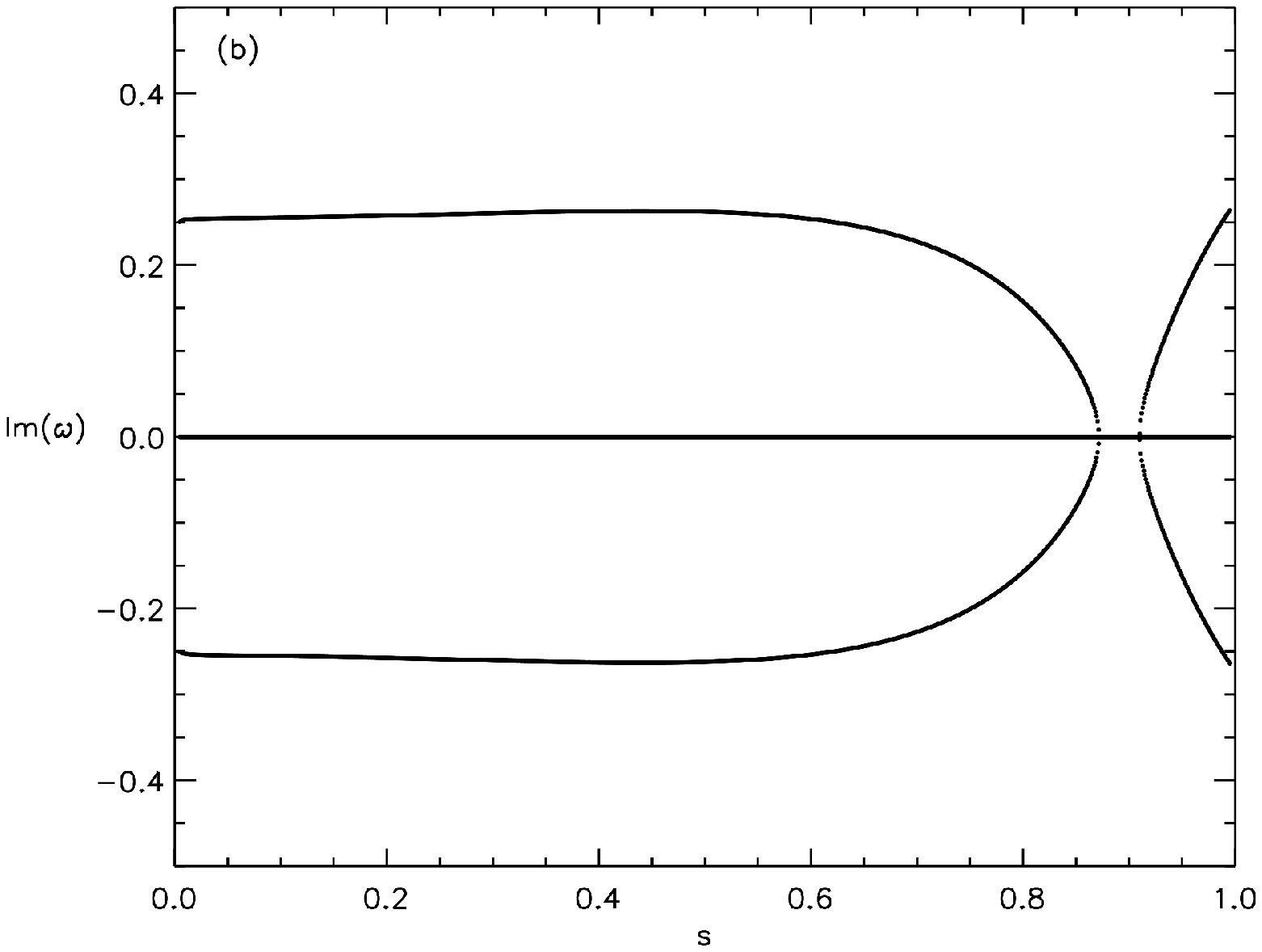}          \\
    \includegraphics[height=0.3\textheight,clip]{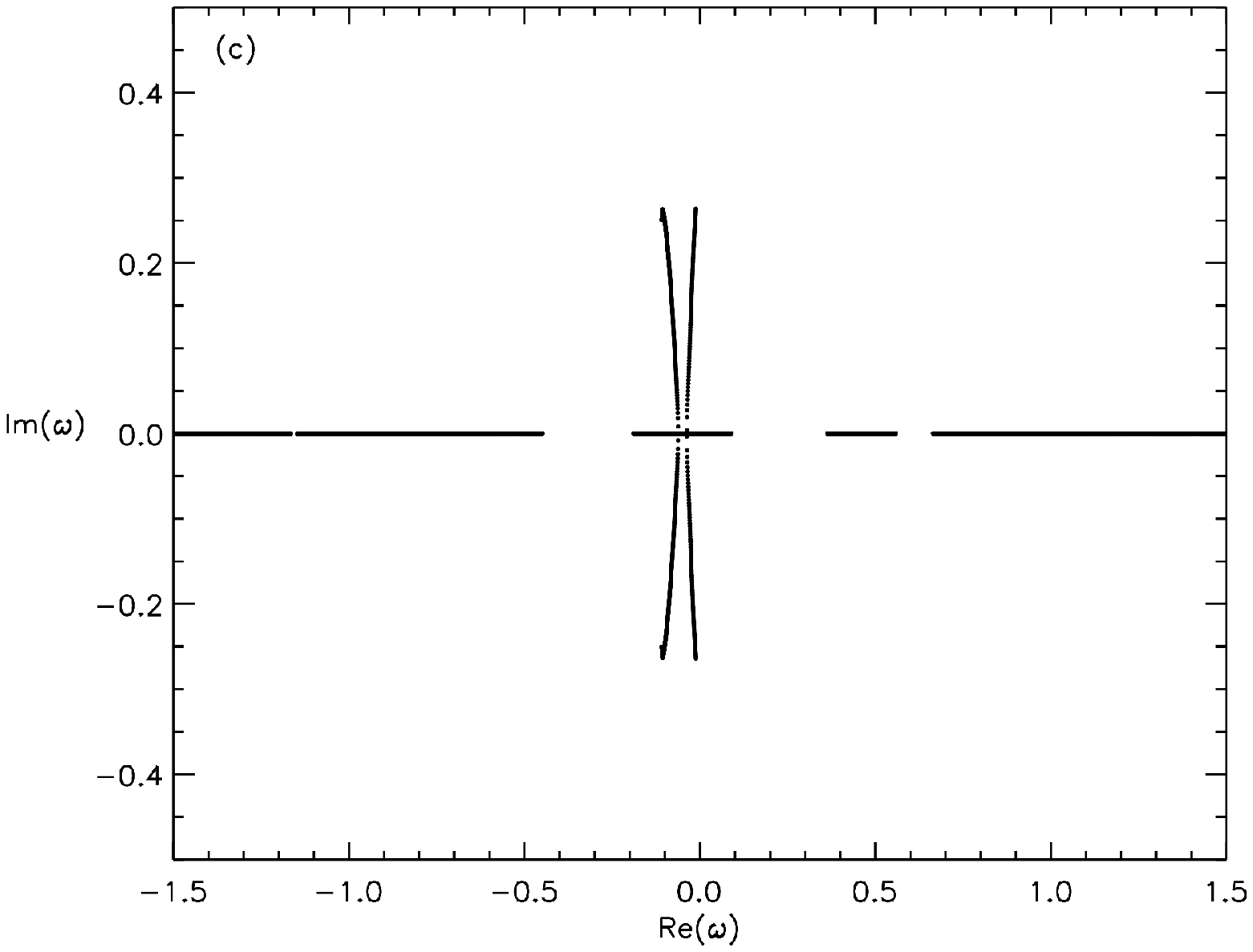} 
  \end{tabular}
  \caption{For the thick disk equilibrium shown in Fig.~\ref{fig:epsilon-0.50-pressure}, (a) the real and (b)
           imaginary parts of the sub--spectrum of the MHD continua as a function of the radial flux 
	   coordinate~$s \equiv \sqrt{\psi}$ and (c) sub--spectrum of the MHD continua in the complex plane are
	   shown for toroidal mode number~$n=-1$ and poloidal mode numbers~$m=[-3,5]$.}
  \label{fig:epsilon-0.50-spectrum}
\end{figure}

For this equilibrium, the continuous MHD spectrum is computed, again for toroidal mode number~$n=-1$ and poloidal 
mode numbers~$m=[-3,5]$. A part of the spectra is plotted in Fig.~\ref{fig:epsilon-0.50-spectrum}. Plot (a) shows 
the real part of the continuous MHD spectrum, while plot (c) shows the imaginary part as a function of the radial 
flux coordinate~$s \equiv \sqrt{\psi}$. Plot (c) contains the eigenfrequencies plotted in the complex plane.  
This shows the existence of overstable modes in a thick accretion disk due to the presence of toroidal flow and 
gravity. If one carefully examines the plot (a) one sees that one of the continuum branches splits in two 
at~$s \approx 0.87$, where the imaginary part becomes zero and at~$s \approx 0.91$ that these two continuum branches 
merge again, whereas the imaginary part becomes non--zero.

\subsection{Maximum growth rate of the MHD continua of thin and thick accretion disks}
\begin{figure}[ht]
  \centering
  \includegraphics[width=0.6\textwidth]{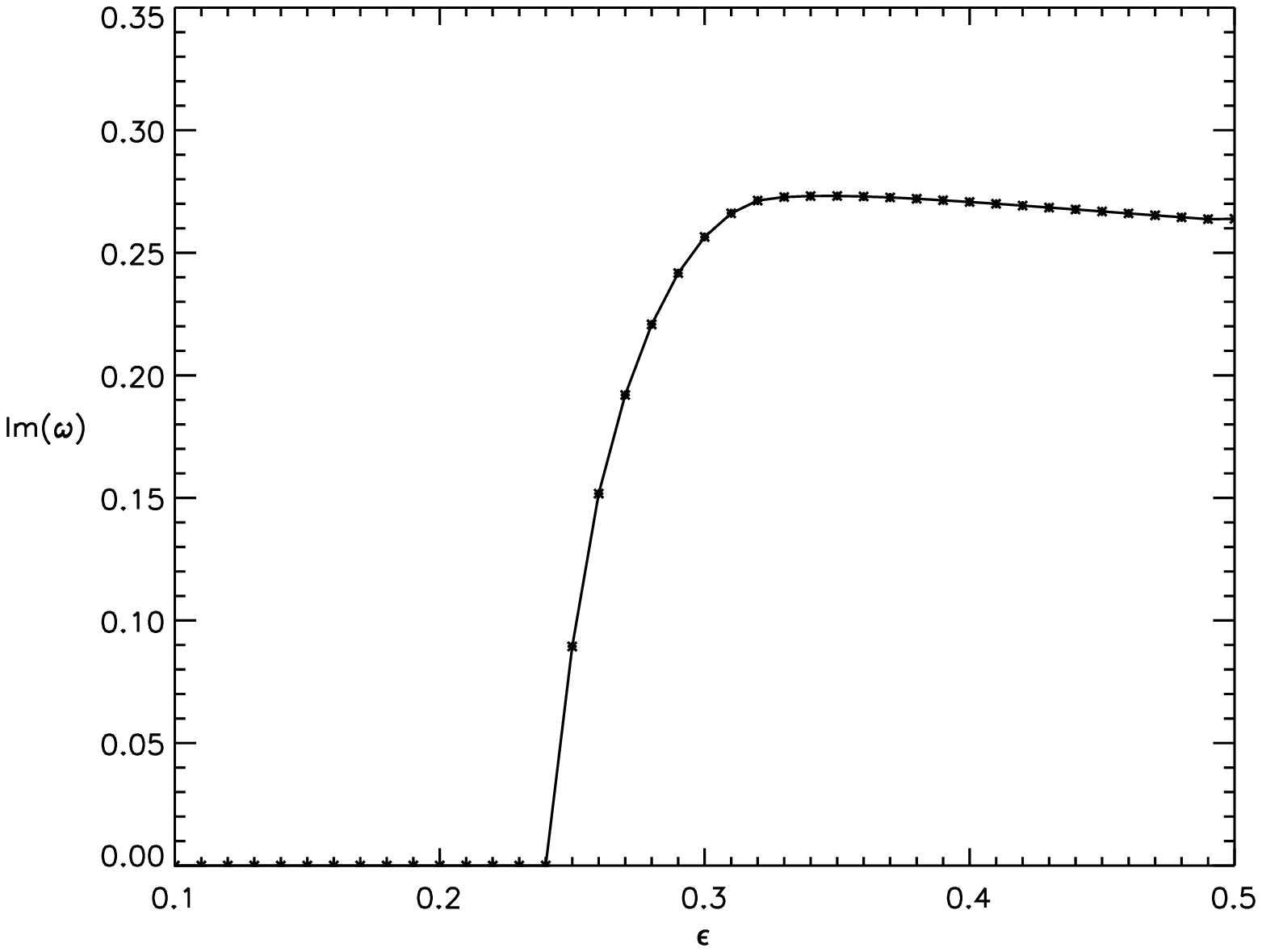}
  \caption{Growth rate of the most overstable continuum mode as a function of the inverse aspect 
           ratio~$\epsilon = a / R_{0}$.}
  \label{fig:growthrate-epsilon}
\end{figure}

The last two cases only differ by the value of the inverse aspect ratio~$\epsilon$. We have systematically 
investigated the influence of this ratio on the existence of overstable modes by varying its value, while keeping 
all other equilibria quantities constant. In this parametric study, we essentially move the disk torus radially inwards
while keeping the radius of the last closed flux surface~$a$ constant. We then gradually evolve from the
equilibrium shown in Fig.~\ref{fig:epsilon-0.10-pressure} ($\epsilon =0.1$) to the one shown in
Fig.~\ref{fig:epsilon-0.50-pressure} ($\epsilon = 0.5$). The growth rate of the most overstable continuum mode is
plotted as a function of the inverse aspect ratio~$\epsilon$ in Fig.~\ref{fig:growthrate-epsilon}. Each point of
the figure represents an equilibrium computed by FINESSE, from which the MHD continua is computed by PHOENIX.
In this figure, one can clearly see that no overstable modes are present in these disks for~$\epsilon=[0.10 , 0.24]$. 
From~$\epsilon=0.24$ up to $0.32$, there is a strong increase of the growth rate of the most overstable mode, 
while from $\epsilon=0.32$ up to at least $0.50$ the growth rate remains approximately the same.

\section{Conclusions \label{sec:conclusions}}
We have presented the equations describing an axisymmetric MHD accretion disk equilibrium and used FINESSE 
to find numerical solutions. We have considered thin as well as thick disks with a plasma beta of the 
order~$0.1$ where the largest part of the disk rotates sub--Keplerian. The numerical solutions show that the pressure 
profile can only decrease in the outward direction if the gravitational potential dominates in the 
pressure equation~\eqref{eq:pressurerho}. 

The stability of the disk equilibria have been analysed theoretically as well as numerically. From theory,
a stability criterion has been derived which closely resembles the Schwarzschild criterion. It is shown that 
in a disk where the density is a flux function, the MHD continua can turn overstable due to the presence of
toroidal rotation and gravity. This instability is not possible for disks where the temperature or the entropy is a 
flux function. The numerical results support our theoretical conclusion. The effect of the gravity on the overstable 
modes can either be stabilizing or destabilizing. In the case of non--constant density the inverse aspect 
ratio~$\epsilon$ has been varied. This study shows that the continua turn overstable if the inverse aspect ratio is 
larger than a certain critical value, approximately 0.24. These localised instabilities will lead to small--scale, 
turbulent dynamics when the equilibria are used in initial conditions for non--linear simulations. This dynamics 
is unrelated the usual magneto--rotational instability. Indeed, these instabilities, as well as the poloidal flow 
driven unstable continuum modes investigated by Goedbloed et al.~\cite{GBHK}, provide a new route to MHD turbulence 
in accretion disks. Furthermore, due to the presence of gravity, it is possible to widen or create new gaps in the 
MHD continua.  It is possible to show this theoretically by expanding the eigenfunctions in poloidal harmonics. 
In these gaps new types of global ideal MHD eigenmodes can exist. The investigation of these global eigenmodes is
left for future work.

\begin{acknowledgements}
This work was carried out within the framework of the European Fusion Programme, supported by the European Communities 
under contract of the Association EURATOM/FOM. Views and opinions expressed herein do not necessarily reflect 
those of the European Commission. This work is part of the research programme of 
the ``Stichting voor Fundamenteel Onderzoek der Materie (FOM)'', which is financially supported by 
the ``Nederlandse Organisatie voor Wetenschappelijk Onderzoek (NWO)''.
\end{acknowledgements}

\end{document}